\documentclass[preprint,review,12pt]{elsarticle}

\usepackage{geometry}
\DeclareUnicodeCharacter{2212}{-}

\usepackage{dsfont}

\usepackage{amsmath}

\usepackage{soul}

\usepackage{float}

\usepackage[toc,page]{appendix}
\newcommand*\xbar[1]{%
   \hbox{%
     \vbox{%
       \hrule height 0.5pt 
       \kern0.5ex
       \hbox{%
         \kern-0.1em
         \ensuremath{#1}%
         \kern-0.1em
       }%
     }%
   }%
}

\usepackage{scalerel}[2014/03/10]
\usepackage{stackengine}
\usepackage{booktabs}
\usepackage{graphicx}

\usepackage{amssymb}
\usepackage{todonotes}
\usepackage{gensymb}
\usepackage{tikz}

\DeclareRobustCommand\full  {\tikz[baseline=-0.6ex]\draw[thick] (0,0)--(0.5,0);}

\DeclareRobustCommand\dotted{\tikz[baseline=-0.6ex]\draw[thick,dotted] (0,0)--(0.54,0);}

\DeclareRobustCommand\chain {\tikz[baseline=-0.6ex]\draw[thick,dash dot dot] (0,0)--(0.5,0);}

\journal{Aerospace Science and Technology}

\begin{document}

\begin{frontmatter}



\title{Large-eddy simulations of the NACA23012 airfoil with laser-scanned ice shapes}

\author[inst1]{Brett Bornhoft}
\author[inst1]{Suhas S. Jain}
\author[inst3]{Konrad Goc}
\author[inst2,inst3]{Sanjeeb T. Bose}
\author[inst1]{Parviz Moin}

\affiliation[inst1]{organization={Center for Turbulence Research, Stanford University},
            city={Stanford},
            postcode={94305}, 
            state={CA},
            country={USA}}

\affiliation[inst3]{organization={ICME, Stanford University},
            city={Stanford},
            postcode={94305}, 
            state={CA},
            country={USA}}

\affiliation[inst2]{organization={Cadence Design Systems},
            city={San Jose},
            postcode={95134}, 
            state={CA},
            country={USA}}
\affiliation[inst3]{organization={The Boeing Company},
    city={Everett},
    postcode={98204},
    state={WA},
    country={USA}}

\begin{abstract}



In this study, five ice shapes generated at NASA Glenn's Icing Research Tunnel (IRT) are simulated at multiple angles of attack \cite{broeren2018icingaeroperf_naca23012}. These geometries target different icing environments, both early-time and longer-duration glaze and rime ice exposure events, including a geometry that results from using a thermal ice-protection system. Using the laser-scanned geometries, detailed representations of the three-dimensional ice geometries are resolved on the grid and simulated using wall-modeled LES. Integrated loads (lift, drag, and moment coefficients) and pressure distributions are compared against experimental measurements in both clean and iced conditions for several angles of attack in both pre-and post-stall regions. The relevant comparisons to the experimental results show that qualitative and acceptable quantitative agreement with the data is observed across all geometries. 

Glaze ice formations exhibit larger and highly nonuniform ice features, such as `horns', in contrast to rime ice formations characterized by smaller, uniformly distributed roughness elements. In wall-modeled LES, it was observed that larger roughness scales in the glaze ice that trigger transition can be accurately resolved. Therefore, it is possible for WMLES to accurately capture the aerodynamics of glaze ice shapes without the need for additional modeling. In contrast, rime ice geometries required additional resolution to accurately represent the aerodynamic loads. This study demonstrates the effectiveness of the wall-modeled LES technique in simulating the complex aerodynamic effects of iced airfoils, providing valuable insights for aircraft design in icing environments and highlighting the importance of accurately representing ice geometries and roughness scales in simulations.
\end{abstract}




\end{frontmatter}



\section{Introduction}
\label{sec:intro}
Predicting the aerodynamic performance of an aircraft in icing conditions is critical, as failures in an aircraft’s ice protection system can compromise flight safety. In response to a series of icing-related accidents, the Federal Aviation Administration (FAA) implemented a new ruling in 2007 that modified the regulations that govern today’s aircraft, which are found in Title 14 Code of Federal Regulations (CFR) Part 25 for Airworthiness Standards: Transport Category Airplanes, by adding a requirement to certify that transport-category airplanes exhibit the same handling/performance in both icing and non-icing conditions \cite{fedregice2007}. A follow-on change to both Parts 25 and 33, adding Appendix O, came into effect in 2015 and focused on extending the new certification requirement to both airframes and engines operating in supercooled large droplet (SLD) and ice crystal icing conditions \cite{fedregice2014}. These rule changes were largely in response to the 1994 accident in Roselawn, Indiana, where an Avions de Transport Régional (ATR) 72 series airplane experienced an un-commanded roll. The National Transportation Safety Board (NTSB) determined that the accident was caused by freezing raindrops that created a spanwise ridge of ice downstream of the wing's deicing boot. 

Prior to the addition of Appendix O, no certification was required for aircraft performance during freezing rain or drizzle events, which induce both SLD and ice crystals. The rule changes have drastically altered how aircraft manufacturers design their airplanes by requiring them to consider the effects of icing in the earliest stages of design. Early designs do not often include ice tunnel testing, primarily due to the cost associated with these tests. Therefore, aircraft designs under these rules rely heavily upon computational fluid dynamics (CFD) simulations. Historically, studying the aerodynamic effects of icing has typically relied on Reynolds-averaged Navier-Stokes (RANS) modeling. RANS models usually struggle to accurately predict the stall behavior, particularly in conditions in which the effect of ice accretion, at early times (less than one minute of accretion), is to only introduce surface roughness to an otherwise smooth surface \cite{rumseyHLCRM2019}. Hence there is a clear need to accurately predict and model ice accretion and its effect on aerodynamics. 

\begin{figure}
    \centering
    \includegraphics[width=1.0\textwidth]{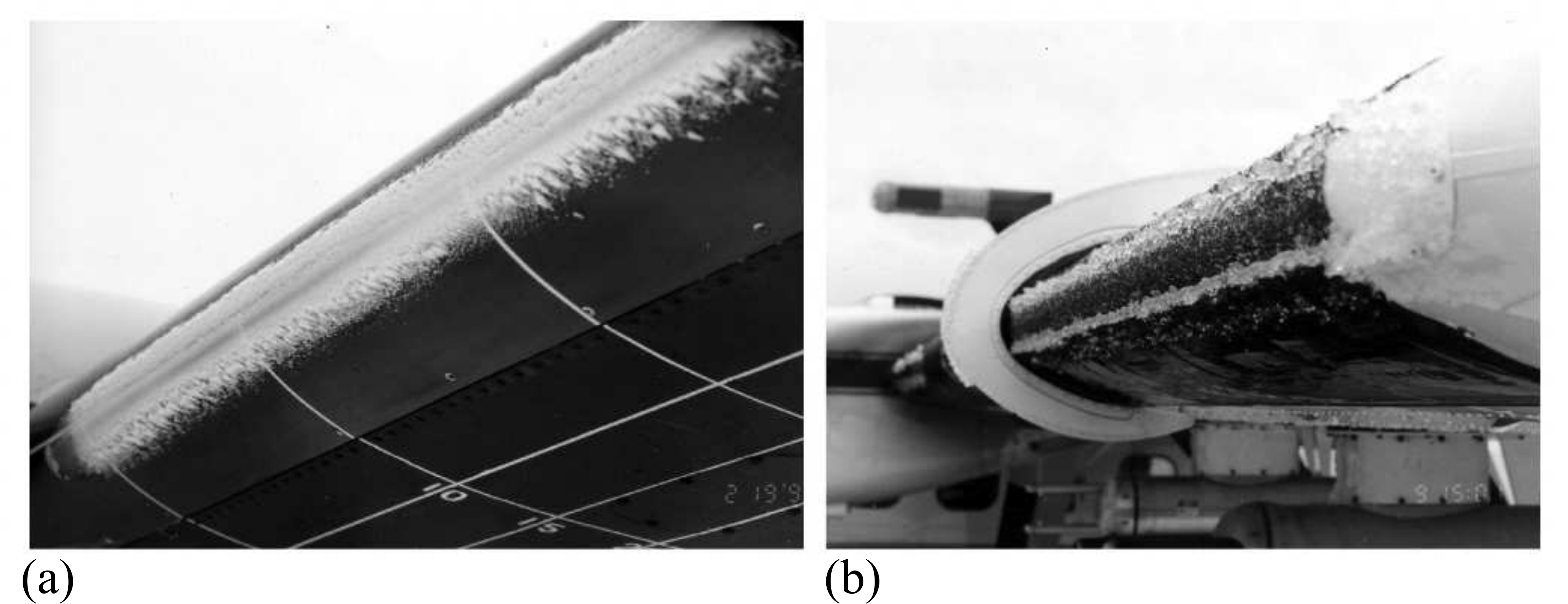}
    \caption{Icing examples from a post-flight test by the NASA Glenn Research Center’s instrumented Twin Otter aircraft: (a) rime ice conditions and (b) glaze ice conditions. \label{fig:icing examples}}
\end{figure}

The two primary types of icing are rime and glaze \cite{gent2000aircraft, politovich2019aircrafticing}. Rime ice typically occurs at low flight temperatures ($<-20^{\circ} $C), low speeds, low atmospheric liquid water concentration, LWC, and small droplet sizes. Rapid and complete freezing of the droplets impacting the aerodynamic surfaces leads to opaque accretion of ice on the wing's leading edge, which is still streamlined. Rime ice's primary aerodynamic impact comes from its characteristic roughness. Glaze ice occurs at higher flight temperatures (near freezing), higher speeds, high atmospheric LWC, and large droplet sizes. Upon impact, the droplets still attach to nucleation points on the wing, but due to larger droplet sizes and higher temperatures, water will run back along the wing body and experience delayed freezing. This results in translucent and smoother ice covering the surface of the aircraft, and when exposed to icing conditions for a longer time, glaze ice leads to so-called horn ice conditions. Ice horns can drastically impact the aerodynamic performance of an aircraft by inducing separation at modest angles of attack, leading to an overall decrease in maximum lift and stall angle \cite{gent2000aircraft}. Figure \ref{fig:icing examples} shows examples of (a) rime and (b) glaze ice from a flight test \cite{politovich2019aircrafticing}.

Encouraged by recent studies using large-eddy simulations (LES) that demonstrate the ability to predict stall characteristics on full aircraft at affordable costs \cite{goc2021large}, we aim to apply this methodology to icing conditions. Measurements of lift, drag, and pitching moment of a NACA23012 airfoil under rime and glaze ice conditions are available at chord-based Reynolds number, $Re_c$, of $1.8$ million (M) in the work of \cite{broeren2018icingaeroperf_naca23012}. Using laser-scanned, detailed representations of the iced geometries from \cite{broeren2018icingaeroperf_naca23012}, we perform LES calculations to assess the aerodynamic impact of the ice in various conditions and the viability of LES as a tool for the simulations of flow over iced aerodynamic surfaces. The simulations are validated by comparing the integrated loads against experimental measurements in both clean and iced conditions at various angles of attack through the onset of stall. 

Most previous studies of icing predictions and its aerodynamic performance typically used lower-fidelity modeling approaches, such as the RANS equations \cite{chi2005comparative, aerospace7040046, potapczuk1991simulation, kwon1997numerical, bragg1991aerodynamic}. More recently, glaze and horn ice shapes have been simulated using a variety of scale-resolving simulation approaches. In \cite{pan2005desice}, a detached-eddy simulation (DES) method was used to study the aerodynamic performance of spanwise ridge ice shapes. Their results showed improved surface pressure distribution compared to RANS methods, but also exhibited worse agreement in the lift coefficient versus angle-of-attack curve. This can be due to the cancellation of errors in the RANS results leading to good agreement for the lift curve, but for incorrect reasons. Other hybrid RANS/LES approaches are used in \cite{duclercq2012zlesice, alam2015dhrlice, zhang2021analysis} where two-dimensional horn and spanwise ridge shapes are simulated and compared with experimental data with marginal improvements in surface pressures and lift coefficients as compared to RANS approaches. A lattice-Boltzmann method (LBM) was used in \cite{konig2015icelbm}, where they simulated glaze and horn ice geometries that are considered in the present study. These simulations marked a first attempt at simulating complex roughness topologies in iced airfoils. We use the results of \cite{konig2015icelbm} as a baseline state-of-the-art literature for the glaze and horn ice geometries and to compare with our simulations. Further discussion of their results are made in Sections \ref{sec:clean}, \ref{sec:glaze}, and \ref{sec:horn}. The wall-modeled LES (WMLES) approaches of \cite{xiao2020icewmles} and \cite{wong2023numerical} focused on large two-dimensional ice shapes. In \cite{xiao2020icewmles}, a NLF-0414/623 airfoil was simulated with a large horn ice shape attached to the leading edge. Both the lift and drag coefficients showed large improvements over RANS methods, where the error in lift was reduced from 19\% to approximately 1\%. In \cite{wong2023numerical}, an immersed boundary method was utilized to simulate the complex horn ice shape of a 2D airfoil. Their wall-modeled approach improved upon the existing unsteady RANS methods, but ultimately underpredicted the lift coefficient in the post-stall region and the critical angle of attack. While the existing literature has predominantly focused on large ice obstructions such as horns and ridges and mostly in 2D airfoils with the extruded ice shape in the third dimension, it is noteworthy that higher-fidelity approaches have not yet been applied to rime ice shapes and airfoils with full 3D ice shape geometries. In light of this gap, our study aims to comprehensively evaluate WMLES for various ice shapes, including large and small ice accretions. We will discuss our findings in the following sections.

In this work, we use WMLES to study the effects of both rime and glaze icing conditions on aerodynamic performance. Five ice shapes are considered: two early-time ice shapes (one each for rime and glaze ice conditions) with accretion times of less than one minute, two long-time ice shapes (horn and streamwise) after five minutes of ice accretion, and one spanwise ridge ice shape mimicking SLD conditions of an airfoil with an ice protection system. At first, ice introduces only roughness to the surface, but after long accretion, large deformations are eventually introduced to the otherwise clean surface. The viability of the WMLES as a tool to predict the effect of icing on aerodynamics in these various conditions is also assessed.

The paper is organized as follows. Section \ref{sec:effect_of_icing} describes the general impact of ice on aerodynamic surfaces. The modeling approach, including details of the LES equations, numerical methods, and wall models, is detailed in Section \ref{sec:modeling_approach}. Section \ref{sec:comp_setup} details the setup for the simulations. A discussion of the computational results is provided in Section \ref{sec:comp_results}. The summary of the study and important conclusions are discussed in Section \ref{sec:conclusion}. A brief discussion on boundary layer transition and its sensitivity to grid resolution in the presence of ice is provided in \ref{sec:app:grid_res}.


\section{Effects of icing: experimental data}\label{sec:effect_of_icing}
NASA Glenn's Icing Research Tunnel (IRT) has conducted research on the effect of ice accretion on aerodynamic bodies since 1944 \citep{potapczuk2013nasaglenn}. The IRT creates a cloud of supercooled water droplets from a series of spray bars that target various droplet diameter ranges, LWC, and temperatures. Recently, laser-scanning techniques have been developed and tested for iced airfoils in the IRT \citep{lee2014icescans}. This advancement provides digital representations of iced airfoils that can be used with rapid prototyping methods (RPM) for manufacturing ice shapes, which can then be used for wind tunnel testing. 

\begin{figure}
    \centering
    \includegraphics[width=0.95\textwidth]{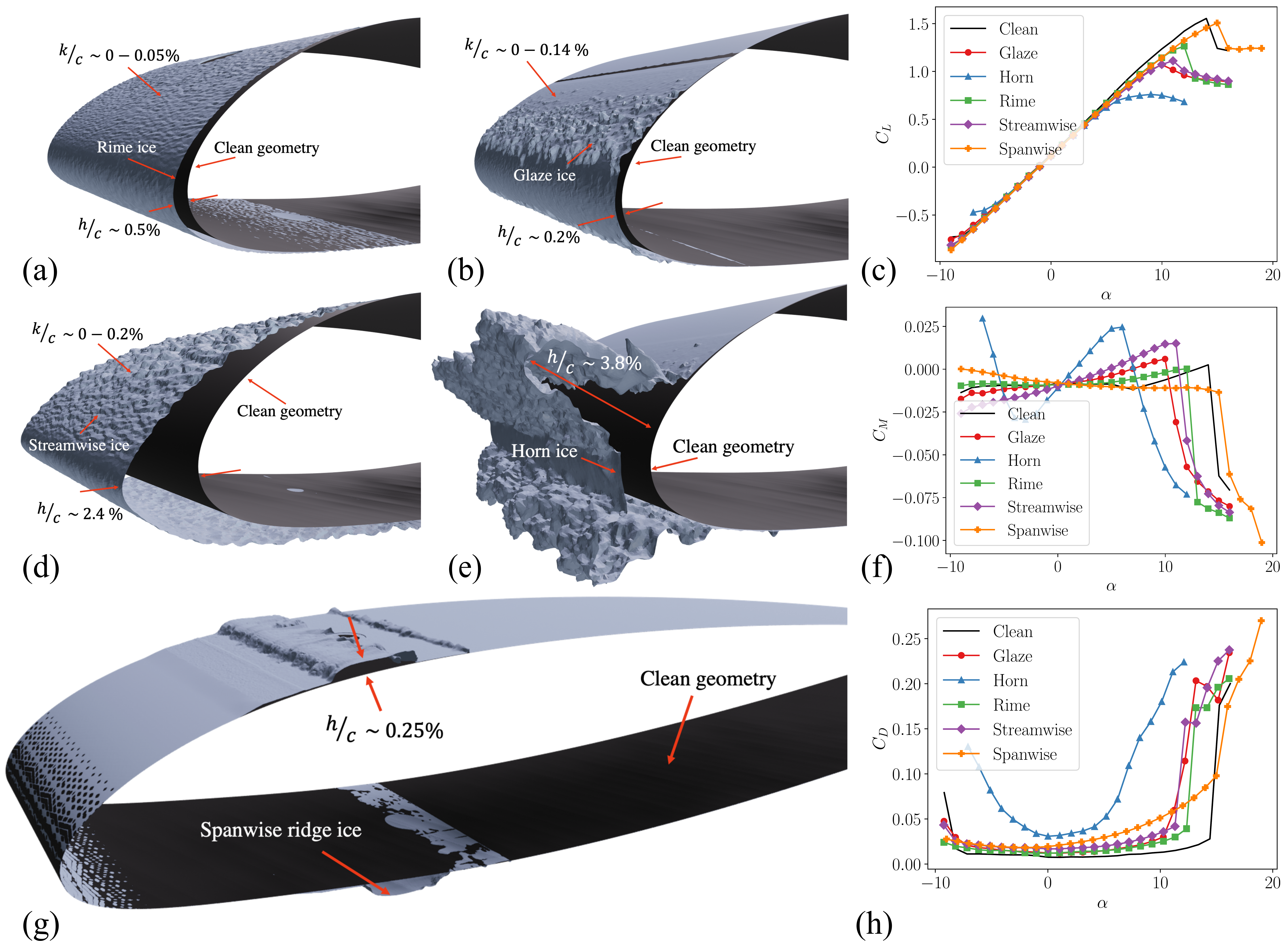}
    \caption{Experimental illustration of the geometric changes to the NACA23012 airfoil due to the (a) early-time rime, (b) early-time glaze, (d) streamwise, (e) horn ice, and (g) spanwise ridge ice accretions on the NACA23012 clean geometry detailing ice scales that geometrically modify the airfoil outer mold line ($h$) and roughness scales ($k$). Aerodynamic performance impacts are shown using the integrated quantities of experimental (c) lift ($C_L$), (f) quarter-chord pitching moment ($C_M$), and (h) wake drag ($C_D$) coefficients for the clean, glaze, horn, rime, streamwise, and spanwise ridge ice airfoils from \cite{broeren2018icingaeroperf_naca23012}.}
    \label{fig:ice performance hit}
\end{figure}

Ice geometries obtained using the laser scanner are shown along with the clean NACA23012 geometry in Figure \ref{fig:ice performance hit}. Here, $k/c$ describes the roughness length scale, $k$, normalized by the clean airfoil chord length, $c$. The value of $h$ is defined as the displacement height due to ice accretion with respect to the clean geometry. The value of $k$ is defined, in this study, as the displacement height minus the average displacement height (averaged along the spanwise direction) and is given by
\begin{equation}
    k(s, z) = h(s, z)-\overline{h(s)}.\label{eq:rough}
\end{equation}
The two length scales represent the bimodal behavior of ice accretion, whereby the ice geometrically modifies the airfoil outer mold line (represented by $h$) and also introduces additional roughness scales (represented by $k$). Details of the ice shapes considered in this study and the specific IRT test conditions are included in Table \ref{tab:ice_description} with specified ice shape, run number, airspeed, mean volumetric droplet diameter, MVD, LWC, total temperature, $T_t$, static temperature, $T_s$, and spray time. Ice accretion tests for rime, glaze, streamwise, and horn ice shapes in Figure \ref{fig:ice performance hit}(a, b, d, e) were conducted at an angle of attack, $\alpha$, of 2\degree. The ice accretion test for the spanwise ridge shape, in Figure \ref{fig:ice performance hit}(g), was done at $\alpha=$ 1\degree.

\begin{table}[]
\resizebox{\textwidth}{!}{%
\begin{tabular}{@{}llllllll@{}}
\toprule
Ice Shape & IRT Run Number & Airspeed (kt) & \begin{tabular}[c]{@{}l@{}}MVD \\ ($\mu m$)\end{tabular} & \begin{tabular}[c]{@{}l@{}}LWC \\ ($g/m^3$)\end{tabular} & $T_t$ ($\degree F/\degree C$) & $T_s$ ($\degree F/\degree C$) & \begin{tabular}[c]{@{}l@{}}Spray time \\ (min)\end{tabular} \\ \midrule
\begin{tabular}[c]{@{}l@{}}Early-time \\ glaze\end{tabular} & ED1974 & 200 & 15 & 0.75 & 28.0/-−2.2 & 18.5/−-7.5 & 0.5 \\
\begin{tabular}[c]{@{}l@{}}Early-time \\ rime\end{tabular} & ED1983 & 200 & 30 & 0.4 & 0.0/-−17.8 & −-9.5/-−23.1 & 1.0 \\
\begin{tabular}[c]{@{}l@{}}Horn \\ \end{tabular} & ED1978 & 200 & 15 & 0.75 & 28.0/-−2.2 & 18.5/-−7.5 & 5 \\
\begin{tabular}[c]{@{}l@{}}Streamwise \\ \end{tabular} & ED1977 & 200 & 30 & 0.4 & 0.0/-−17.8 & -−9.5/-−23.1 & 5 \\
\begin{tabular}[c]{@{}l@{}}Spanwise\\ Ridge\end{tabular} & ED1967 & 175 & 15 & 0.64 & 24.0/−-4.4 & 16.8/−-8.4 & 9.5 \\ \bottomrule
\end{tabular}
}
\caption{Details of the ice accretion parameters for the ice shapes generated in the NASA Glenn IRT \cite{broeren2018icingaeroperf_naca23012}.}
\label{tab:ice_description}
\end{table}
Figure \ref{fig:ice performance hit} shows the (c) lift, $C_L$, (f) quarter-chord pitching moment, $C_M$, and (h) wake drag, $C_D$, coefficients for the clean and the five ice geometries obtained from the experimental results of \cite{broeren2018icingaeroperf_naca23012}. 
The rime, glaze, streamwise, and horn geometries result in an early onset of stall accompanied by a significant rise in drag and an increased nose-down pitching moment. The spanwise ridge geometry results in a delayed stall angle due to the local energization of the boundary layer generated from a small separation bubble downstream of the ridge \cite{whalen2006considerations}. This phenomenon is discussed at length in Section \ref{sec:spanwise_ridge}. Similar to the clean airfoil, the rime and spanwise ridge ice geometries have aggressive stall behavior with a sharp decrease in $C_L$. By contrast, the glaze, horn, and streamwise ice geometries have a shallower stall behavior with an earlier onset of stall. For the two early-time ice geometries, the scales of ice are less than $1\%$ of the chord length, yet they still result in a drastic reduction in the aerodynamic performance. The horn ice geometry yields the largest aerodynamic effect, given its larger and irregular shape, with a region of a nose-up pitching moment at moderate angles of attack ($\alpha \approx 4\degree-6\degree$).

\section{Modeling approach}\label{sec:modeling_approach}

\subsection{LES equations}
We utilize LES to simulate the flow about various ice shapes. These methods rely on resolving the large scales and providing closure models for turbulence in the sub-grid regime. We solve the governing equations for a low-pass filtered, compressible Navier–Stokes system for mass, momentum, and total energy, which can be written as
\begin{equation}
\frac{\partial \bar{\rho}}{\partial t} + \frac{\partial \widetilde{u}_i \bar{\rho}}{\partial x_i} = 0,
\label{eq:ass}
\end{equation}
\begin{equation}
\frac{\partial \bar{\rho}\widetilde{u}_i}{\partial t} + \frac{\partial \bar{\rho} \widetilde{u}_i \widetilde{u}_j}{\partial x_j} = -\frac{\partial \bar{p}}{\partial x_i} +\frac{\partial \widetilde{\tau}_{ij}}{\partial x_j} -\frac{\partial \tau_{ij}^{sgs}}{\partial x_j},
\label{eq:mom}
\end{equation}
\begin{equation}
\frac{\partial \bar{E}}{\partial t} + 
\frac{\partial \widetilde{u}_i \bar{E}}{\partial x_i} + 
\frac{\partial \widetilde{u}_i \bar{p}}{\partial x_i} = 
\frac{\partial \widetilde{\tau}_{ij} \widetilde{u}_i}{\partial x_j} -\frac{\partial \tau_{ij}^{sgs} \widetilde{u}_i}{\partial x_j} + 
\frac{\partial }{\partial x_i}\left(\lambda \frac{\partial \bar{T}}{\partial x_i} \right)
- \frac{\partial Q_i^{sgs}}{\partial x_i},
\label{eq:energy}
\end{equation}
where $\widetilde{\cdot}$ and $\bar{\cdot}$ represent Favre--and Reynolds-averaged quantities, respectively, $\rho$ is the density, $\vec{u}$ is the velocity, $p$ is the pressure, $\widetilde{\tau}_{ij} = 2\mu(\widetilde{T})\widetilde{\mathbb{D}}_{ij} - 2\mu(\widetilde{T})({\partial \widetilde{u}_k}/{\partial x_k})\delta_{ij}/3$ is the resolved Cauchy stress tensor, $\mu(\widetilde{T})$ is the dynamic viscosity, $\widetilde{\mathbb{D}}_{ij}=\{({\partial \widetilde{u}_i}/{\partial x_j}) + ({\partial \widetilde{u}_j}/{\partial x_i})\}/2$ is the resolved strain-rate tensor, $\widetilde{E}=\bar{\rho}(\widetilde{e} + \widetilde{u}_i\widetilde{u}_i/2)$ is the resolved total energy per unit volume, $\lambda$ is the thermal conductivity, and $T$ is the temperature. The subgrid terms, denoted by superscript $sgs$, are the terms that account for the effect of subgrid stress and heat flux on the resolved scales and are defined as
\begin{equation}
\begin{split}
    \tau_{ij}^{sgs} = \bar{\rho}(\widetilde{{u}_i {u}_j} - \widetilde{u}_i \widetilde{u}_j),\\
    \vec{Q}_{i}^{sgs} = \bar{\rho}(\widetilde{e u_i} - \widetilde{e} \widetilde{u_i}).
\end{split}
\end{equation}
The subgrid closures are modeled using the dynamic-Smagorinsky approach of \cite{moin1991dynamic}.

\subsection{Numerical modeling}
The solver employed is charLES \citep{goc2021large}, a massively parallel, second-order, low-dissipation finite-volume solver.
The numerical scheme is based on discretely kinetic energy--and entropy-preserving formulation \citep{tadmor2003entropy,honein2004higher,chandrashekar2013kinetic} that has been shown to be suitable for coarsely resolved LES of turbulent flows that are especially sensitive to numerical dissipation. The discretization is suitable for arbitrary unstructured, polyhedral meshes, and the solutions contained herein are computed on unstructured grids constructed based on Voronoi diagrams. The use of Voronoi diagram-based meshes allows for the rapid and automatic generation of high-quality grids with some guaranteed properties (for instance, the vector between two adjacent Voronoi sites is parallel to the normal of the face that they share). Moreover, it is suitable for complex geometries because it results in body-fitted mesh without the need for manual intervention, including in situations with the presence of realistic rough walls like in the current study. Figure \ref{fig:wall_model_schematic} shows a zoomed-in view of an LES Voronoi grid adjacent to a rough wall.  The time advancement is performed using a three-stage explicit Runge Kutta scheme \citep{gottlieb2001strong}, and the spatial discretization is formally second-order accurate. Additional details on the numerical discretization and the grid generation can be found in \cite{fu2021}, \cite{lozano2020prediction}, and \cite{bres2018}. 

\subsection{Wall modeling}

\begin{figure}
    \centering
    \includegraphics[width=\textwidth]{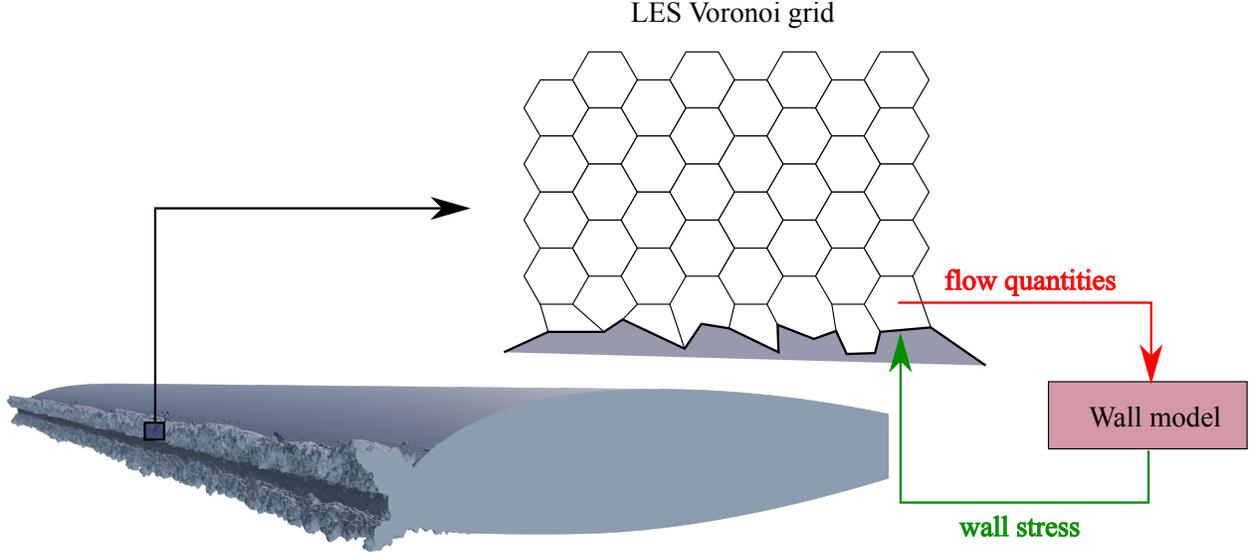}
    \caption{Schematic representing the wall modeling procedure.}
    \label{fig:wall_model_schematic}
\end{figure}

In a wall-modeled LES, only turbulent eddies that scale with the boundary layer thickness are resolved and the remaining effect of the near-wall region is modeled, resulting in significant computational cost savings. In this work, near-wall regions are modeled using an equilibrium wall-modeling approach in which we assume the pressure gradient balances the advective terms, which leads to the constant stress layer approximation \citep{bose2018wmles},
\begin{equation}
    \frac{d \tau}{d n} = \frac{d}{dn} \left(\mu \frac{d U}{d n} - \xbar{\rho u' v'} \right) = 0,
\end{equation}
which can be solved for the wall stress, given the no-slip condition at the wall, a closure for Reynolds shear stress, and the velocity at the matching first grid point from the LES solution. The schematic in Figure \ref{fig:wall_model_schematic} illustrates this procedure.
Additional details on equilibrium wall modeling can be found in \cite{goc2021large}.

\section{Computational setup}\label{sec:comp_setup}
\begin{figure}
    \centering
    \includegraphics[width=0.75\textwidth]{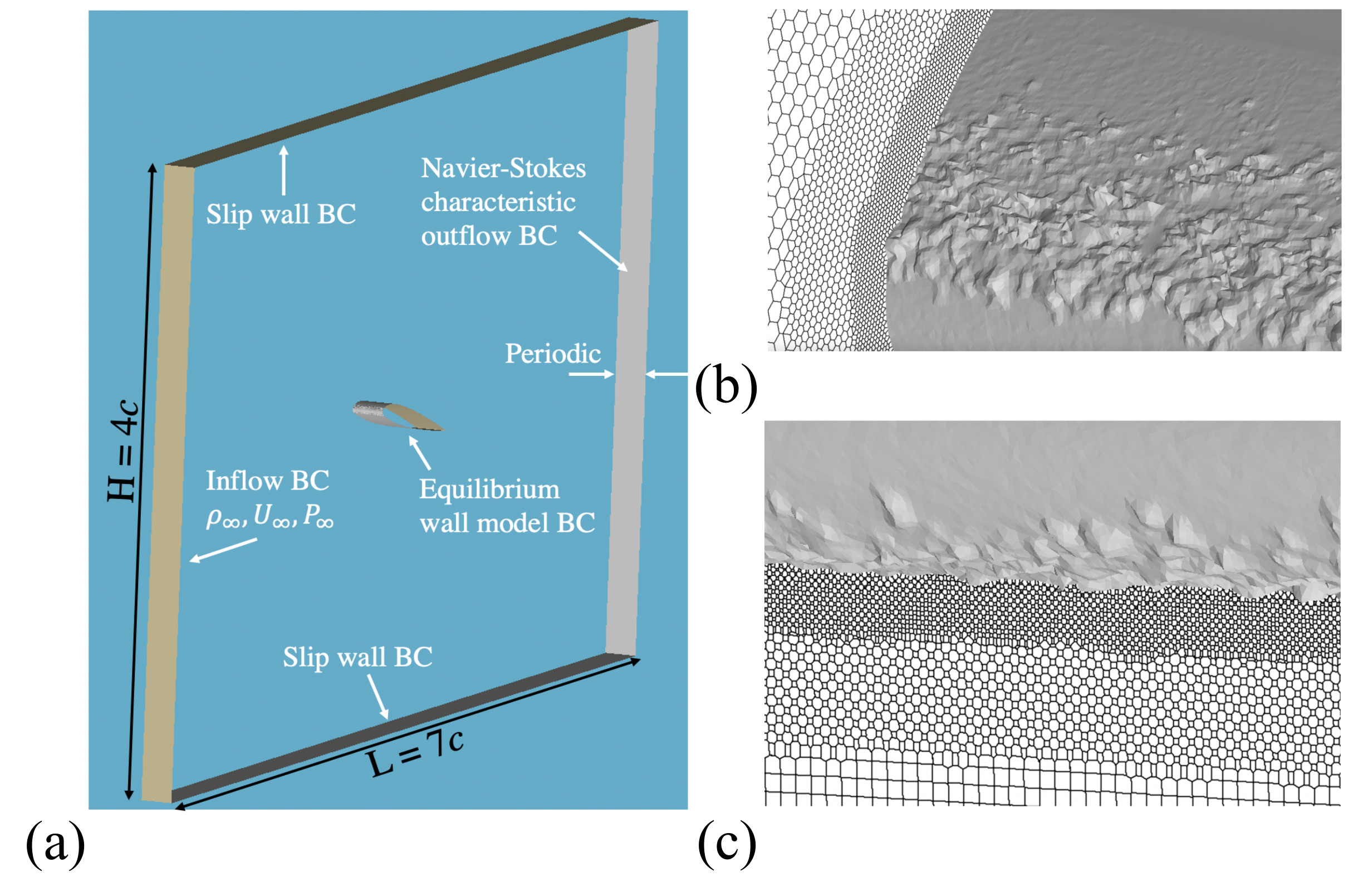}
    \caption{(a) Schematic showing the computational domain setup highlighting the inflow boundary condition (BC), outflow BC, equillibrium wall model BC (applied on the airfoil surface), slip wall BC for the top and bottom of the domain, and periodic BC in the spanwise direction. (b,c) HCP element slices highlighting the resolution near rough surfaces: (b) streamwise and (c) spanwise slices. Lengths are defined with respect to the chord length ($c$).}
    \label{fig:domain_description}
\end{figure}

The cases simulated utilize the NACA23012 two-dimensional airfoil. The flow field is characterized by a free-stream Mach number, $M_\infty$, of $0.18$ and a chord-based Reynolds number, $Re_c$, of $1.8$ M. As highlighted in Section \ref{sec:effect_of_icing}, six geometries are considered in this study; the clean NACA23012 geometry, two glaze ice geometries (early-time and horn), two rime ice geometries (early-time and streamwise), and the spanwise ridge ice case mimicking SLD conditions with an ice protection system. Hereafter, we refer to the cases as clean, glaze ice, horn ice, rime ice, streamwise ice, and spanwise ridge ice, respectively. A set of angles of attack ($\alpha$) are simulated by rotating the airfoil about the quarter chord location. The geometries are modeled in a rectangular domain, as shown in Figure \ref{fig:domain_description}(a), approximately mimicking the wind tunnel effects. Inflow conditions are specified using free-stream pressure, density, and Mach number ($P_\infty=101,325$ Pa, $\rho_\infty=1.225$ kg/m$^2$, and $M_\infty=0.18$). The outflow boundary is modeled as the non-reflecting characteristic boundary condition \cite{nscbc_poinsot}. The top and bottom of the domain are modeled using an inviscid wall boundary conditions. An algebraic equilibrium wall model is applied at the airfoil surfaces \cite{lehmkuhl2016flow}. The spanwise boundaries are treated as periodic conditions. Periodicity is forced on the streamwise, horn, and spanwise ridge ice cases by mirroring a spanwise section of the domain. The clean, rime, and glaze ice geometries do not require mirroring as the deviation in the spanwise boundaries is minimal and periodicity can be enforced. Statistics are gathered throughout the simulation and averaged across at least $20$ chord-based flow-through times. 

\begin{figure}
    \centering
    \includegraphics[width=\textwidth]{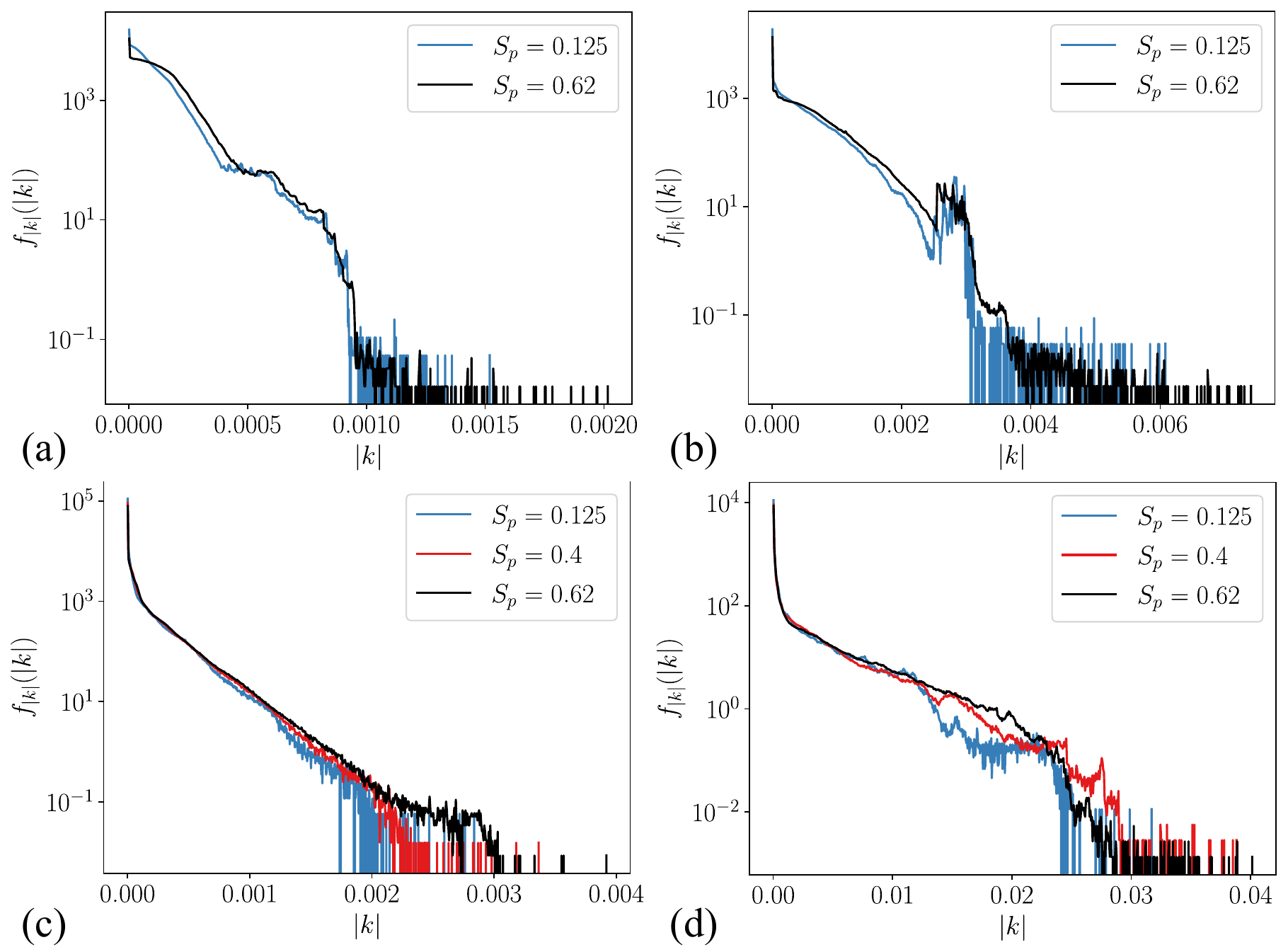}
    \caption{Normalized PDF of $|k|$ with simulated spanwise extents ($S_p$) for the (a) early-time rime, (b) streamwise, (c) early-time glaze, and (d) horn ice shapes. Here, $k$ and $S_p$ are normalized by $c$.}
    \label{fig:pdf_rough}
\end{figure}

Details of the specific spanwise extents, $S_p$, are outlined in Table \ref{tab:grid_description}. Only a portion of the ice geometry is used along the spanwise direction for each simulation. The experimental ice accretion test has a span of $0.62c$. For the subsequent wind tunnel tests, this section was copied three times to fit the wind tunnel width, resulting in a total span of $1.86c$ \cite{broeren2018icingaeroperf_naca23012}. Including the entire span becomes computationally prohibitive, and therefore, we here choose to simulate only a portion of each geometry along the spanwise direction and assume that the chosen portion is representative of the full ice geometry. To determine the effect of this assumption, we plot the probability density function (PDF) of the magnitude of the roughness length scale, $f_{|k|}(|k|)$, in Figure \ref{fig:pdf_rough} for the selected subsection of the ice shapes and its comparison with the full geometry. The early and long-time exposure to rime ice conditions results in nearly homogeneous spatial distribution in the spanwise direction. Therefore, we see similar PDF profiles for the subset and full geometric representation of the ice in Figure \ref{fig:pdf_rough} (a) and (b). For the glaze ice (early and long-time exposure) geometries [Figure \ref{fig:pdf_rough} (c) and (d)], we observe larger deviations between the full and subset ice geometries, particularly for the long-time exposure case. This is due to the larger spanwise variations observed in horn-ice geometries. To determine sensitivities to this variation, we perform additional spanwise extent studies for both glaze ice geometries (see, Table \ref{tab:grid_description}).

Water-tight geometries of the airfoils are used as the surfaces for generating a Voronoi diagram following the algorithm proposed by \cite{du2006convergence}. The complex roughness elements present in the ice are resolved using the Voronoi points. This leads to a series of body-fitted unstructured meshes. Figure \ref{fig:domain_description} (b) and (c) illustrate the ability of the Voronoi points to resolve the complex roughness elements present in the ice. The coarsest grid for the clean ice geometry contains $\approx1.2$ M control volumes (CV). Each successive grid is obtained by homothetically refining the near-wall cells by a factor of two. Each additional layer adds ten cells in the wall-normal direction and isotropically refines the elements in the other two directions. This results in $\approx 4$ M CV for the medium clean case, and similarly $\approx14.25$ M CV for the fine, and $\approx54$ M CV for the extra fine cases. Approximate grid cell counts, $S_p$, and number of points per trailing edge boundary layer height, $\delta/\Delta_{min}$, are detailed in Table \ref{tab:grid_description}. Here, $\delta$ is the boundary layer thickness, and $\Delta_{min}$ is the minimum grid length scale in the mesh. An additional level of refinement is added for the first quarter of the chord, as it was observed that leading edge refinement contributed to capturing the correct flow acceleration. This is essential in predicting stall, especially at higher angles of attack \cite{goc2023thesis}. Additional refinement is also applied to all rough surfaces on the airfoil's upper and lower surfaces.

\begin{table}[]
\centering
\begin{tabular}{@{}lllll@{}}
\toprule
Ice Shape & Ref. Level & \begin{tabular}[c]{@{}l@{}}Cell count\\ (M CV)\end{tabular} & $S_p$ &  $\delta/\Delta_{min}$ \\ \midrule
Clean & \begin{tabular}[c]{@{}l@{}}Coarse\\ Medium\\ Fine\\ Extra fine\end{tabular} & \begin{tabular}[c]{@{}l@{}}1.2\\ 4\\ 14.25\\ 54\end{tabular} & 0.125 & \begin{tabular}[c]{@{}l@{}}8\\ 14\\ 25\\ 40\end{tabular} \\ \midrule
\begin{tabular}[c]{@{}l@{}}Early-time\\ glaze\end{tabular} & \begin{tabular}[c]{@{}l@{}}Coarse\\ Medium\\ Fine\\ Fine\\ Fine\\ Fine\end{tabular} & \begin{tabular}[c]{@{}l@{}}2\\ 7.15\\ 27\\ 43\\ 85\\ 170\end{tabular} & \begin{tabular}[c]{@{}l@{}}0.2\\ 0.2\\ 0.2\\ 0.4\\ 0.8\\ 1.6\end{tabular} & \begin{tabular}[c]{@{}l@{}}8\\ 14\\ 25\\ 25\\ 25\\ 25\end{tabular} \\ \midrule
\begin{tabular}[c]{@{}l@{}}Early-time\\ rime\end{tabular} & \begin{tabular}[c]{@{}l@{}}Fine\\ Extra fine\end{tabular} & \begin{tabular}[c]{@{}l@{}}21\\ 109\end{tabular} & \begin{tabular}[c]{@{}l@{}}0.125\\ 0.125\end{tabular} & \begin{tabular}[c]{@{}l@{}}25\\ 40\end{tabular} \\ \midrule
Horn & \begin{tabular}[c]{@{}l@{}}Fine\\ Fine\end{tabular} & \begin{tabular}[c]{@{}l@{}}33.25\\ 110\end{tabular} & \begin{tabular}[c]{@{}l@{}}0.25\\ 0.8\end{tabular} & \begin{tabular}[c]{@{}l@{}}25\\ 25\end{tabular} \\ \midrule
Streamwise & Fine & 34 & 0.25 & 25 \\ \midrule
\begin{tabular}[c]{@{}l@{}}Spanwise\\ Ridge\end{tabular} & Fine & 37 & 0.25 & 25 \\ \bottomrule
\end{tabular}
\caption{Grid refinement details for the different geometries comparing refinement levels; cell counts in millions of control volumes (M CV); chord normalized spanwise extent ($S_p$); and points per boundary layer thickness ($\delta$), where $\Delta_{min}$ is the minimum grid length scale. Note: All values of $\delta/\Delta_{min}$ are referenced with respect to the clean ice simulation's boundary layer height at the trailing edge of the airfoil.}
\label{tab:grid_description}
\end{table}

\section{Computational Results}\label{sec:comp_results}
We compare with the experimental data of \cite{broeren2018icingaeroperf_naca23012} in each case. We use the uncorrected force balance measurements for comparison based on a domain size sensitivity study. These simulations showed that the chosen domain is more representative of a wall-bounded wind tunnel rather than a free-air configuration. In addition to force balance measurements, a wake rake determines momentum losses across the airfoil. We include these measurements in the drag coefficient comparisons of Figure \ref{fig:fig3_clean_perf}(b) \citep{monastero2014validation}. Previously \cite{konig2015icelbm} used an LBM to simulate the clean, glaze, and horn ice geometries used in the current study. To the best of our knowledge, no other studies consider these ice geometries other than the present work and the LBM results in \cite{konig2015icelbm}. Where appropriate, we have included the LBM results in the plots throughout this work.

\subsection{NACA23012 clean airfoil}\label{sec:clean}

\begin{figure}
    \centering
    \includegraphics[width=1.0\textwidth]{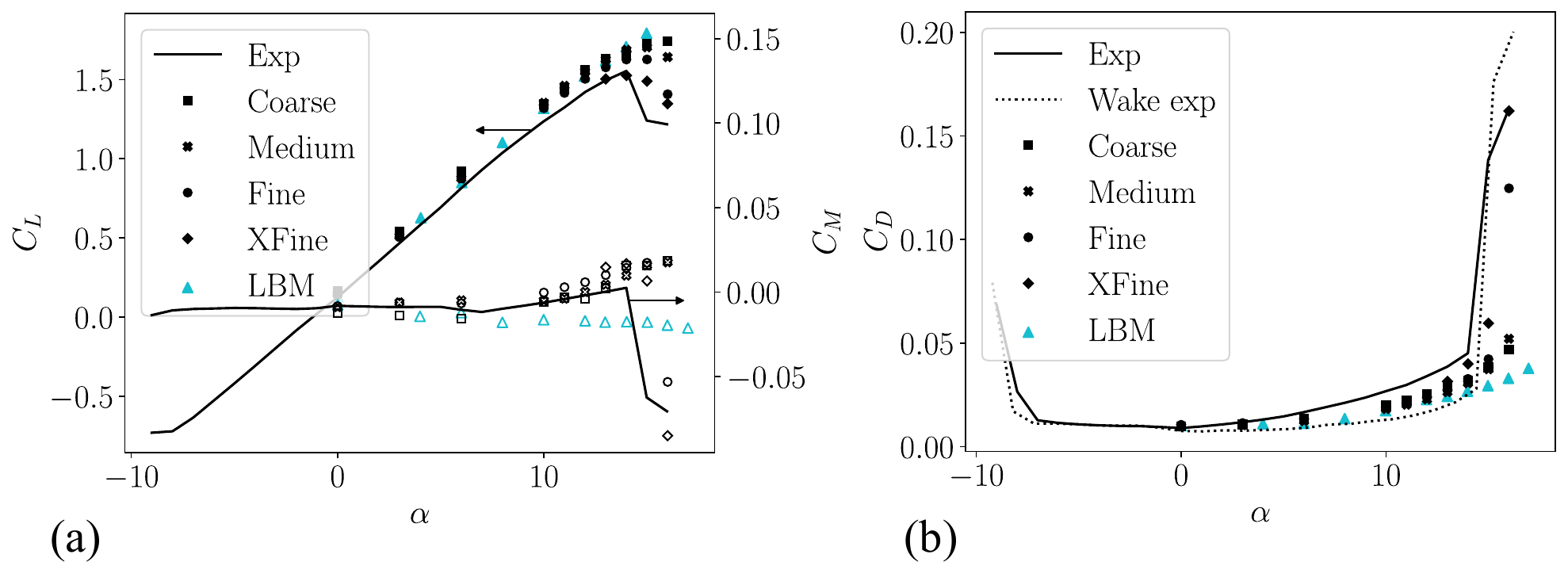}
    \caption{Comparing (a) lift ($C_L$, closed symbols), moment ($C_M$, open symbols), (b) and drag ($C_D$) coefficients of the present WMLES results at three grid resolutions to experimental wake (\dotted) and force balance (\full) measurements \citep{broeren2018icingaeroperf_naca23012} as well as LBM results \citep{konig2015icelbm} for the clean NACA23012 geometry as a function of angle of attack ($\alpha$).}
    \label{fig:fig3_clean_perf}
\end{figure}

Accurate prediction of lift, drag, and moment coefficients for the clean airfoil geometry are necessary to establish simulation credibility before simulating the airfoil using various ice geometries. In Figure \ref{fig:fig3_clean_perf}, aerodynamic coefficients are plotted as a function of the angle of attack for a series of grid resolutions. As the angle of attack increases, the lift coefficient linearly increases while the moment and drag coefficients remain approximately constant, with minor increases nearing the critical stall angle. Around $\alpha=14^{\degree}$, a sharp decline in lift occurs due to the rapid onset of stall. This is accompanied by a nose-down pitching moment and a rapid rise in drag. We note that the LBM results do not show the stall that is seen in the experiments of \cite{broeren2018icingaeroperf_naca23012}. König \textit{et al.} attribute this to not correctly reproducing a laminar separation bubble at the leading edge of the airfoil \cite{konig2015icelbm}. According to Broeren \textit{et al.}, the NACA 23012 airfoil exhibits a leading-edge stall type where the flow abruptly separates due to the bursting of a small leading-edge laminar separation bubble \cite{broeren2011nasatr}. While the WMLES solution also does not resolve a laminar separation bubble, the results presented here do show stall behavior starting with the medium grid resolution\footnote{In \cite{konig2015icelbm}, no resolution is reported for the clean geometry. The early-time glaze ice geometry results in $\approx 11$ MCV for an equivalent $S_p=0.2$. This lies between our medium and fine resolutions for the early-time glaze ice geometry.} Upon further refinement, the stall is more representative of the experimental data, albeit not as abrupt. In the linear region of the lift curve, both methods agree with the slope of the lift curve. An extra-fine resolution is simulated at higher angles of attack. Here, the additional refinement leads to a good agreement in the regime just prior to stall. A general improvement is observed compared to the previous LBM results across all quantities. The overprediction of a nose-down pitching moment at finer angles of attack is discussed in Section \ref{sec:span}, relating the phenomenon to spanwise periodic restrictions. For this study, the results of the clean airfoil can be considered acceptable, as they have improved upon previous results in the literature and capture the experimental stall behavior. In the rest of the paper, we report the fine grid results for the clean case when comparing them to other iced cases. We choose this resolution because it is considered tractable for simulating more complicated configurations of interest, such as swept wings, engine nacelles, and the high-lift common research model, under icing conditions \citep{goc2021large}.  

\subsection{Early-time glaze ice geometry}\label{sec:glaze}

\begin{figure}
    \centering
    \includegraphics[width=1.0\textwidth]{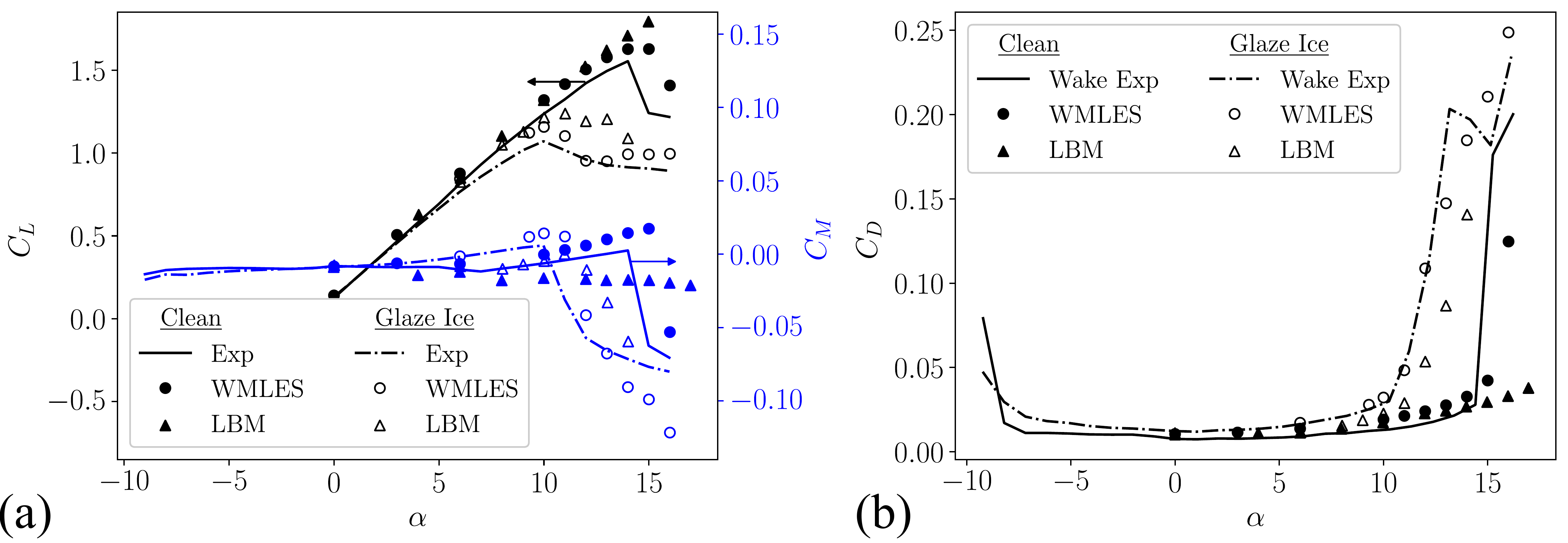}
    \caption{(a) Lift ($C_L$), moment ($C_M$), and (b) drag ($C_D$) coefficients comparing the clean (closed symbols, \full) and glaze ice (open symbols, \chain) geometries of the experimental \citep{broeren2018icingaeroperf_naca23012}, LBM \citep{konig2015icelbm}, and present WMLES results.}
    \label{fig:ed1974_perf}
\end{figure}

\begin{figure}
    \centering
    \includegraphics[width=1.0\textwidth]{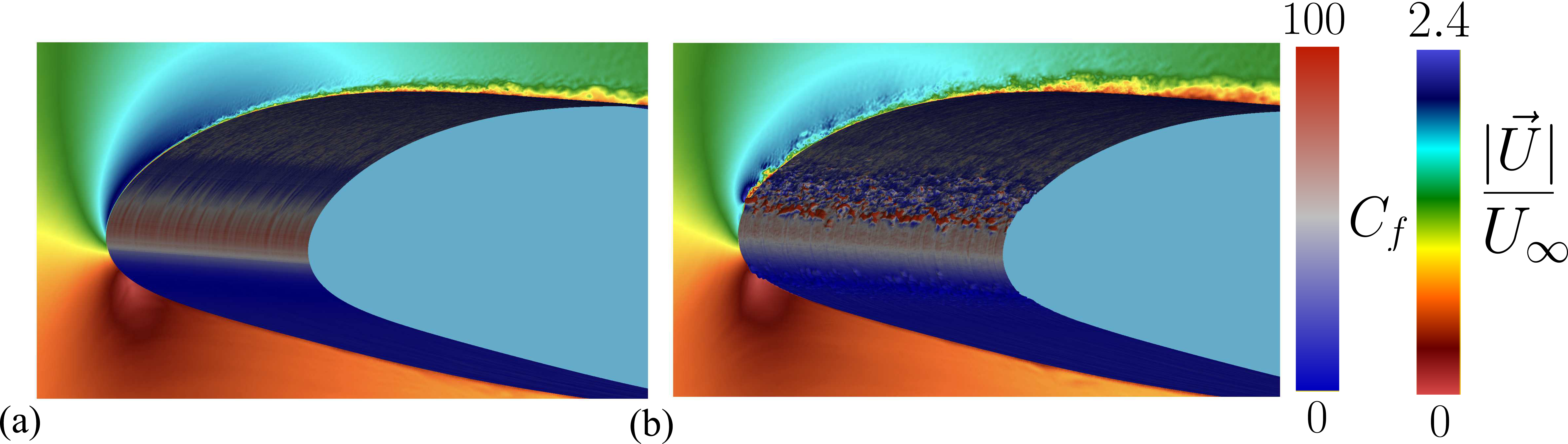}

    \caption{Center-line flow-field slice colored by velocity magnitudes with surface plot colored by wall shear stress of (a) NACA23012 clean geometry and (b) early glaze ice geometry, both at $\alpha=9.3^{\degree}$.}
    \label{fig:clean_ed1974_images}
\end{figure}

\begin{figure}
    \centering
    \includegraphics[width=0.99\textwidth]{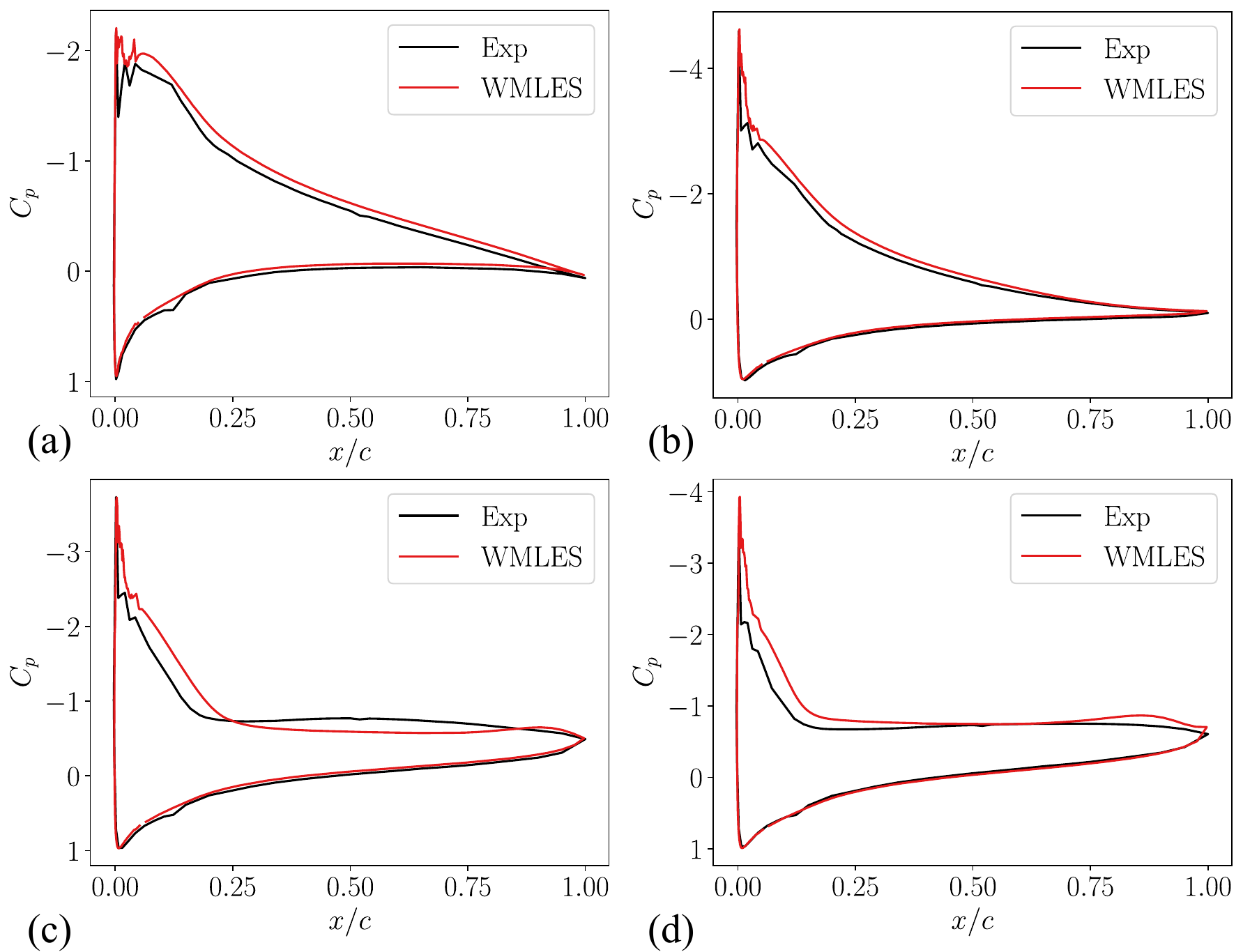}
    \caption{Pressure coefficient ($C_p$) comparison to experimental data for the glaze ice geometry with angles of attack at (a) $6\degree$, (b) $10\degree$, (c) $12\degree$, and (d) $14\degree$ \citep{broeren2018icingaeroperf_naca23012}.}
    \label{fig:ed1974_cp}
\end{figure}

The first ice geometry considered in this work is an early-time glaze ice case, shown in Figure \ref{fig:ice performance hit}(b). The glaze ice geometry has a smoother ice section near the leading edge, with roughness elements downstream on both the upper and lower wing surfaces (see Figure \ref{fig:roughness}(g), (h), and (i)). Prolonged exposure to icing conditions for this geometry leads to ice horns. As stated in Section \ref{sec:clean}, we use the fine grid as the benchmark grid resolution. Table \ref{tab:grid_description} has a total grid count of 27 M CV for the glaze ice case. Similar to the clean NACA23012 geometry, we compare with the experimental results and the LBM results of \cite{konig2015icelbm} in Figure \ref{fig:ed1974_perf}. Improved results are observed compared to those of \cite{konig2015icelbm} for the lift and drag coefficients in the present study. Here, the stall is less pronounced than that of the clean airfoil. A rapid transition can be observed in the comparison of velocity magnitude slices with wall-shear-stress colored surfaces in Figure \ref{fig:clean_ed1974_images}. Compared to the smooth streaks in the clean geometry, the glaze ice geometry's large rough elements immediately trigger boundary layer transition (see \ref{sec:app:grid_res}). The WMLES results predict the stall angle of attack correctly (after $\alpha=10\degree$), while the LBM method predicts stall at one degree past the experimental measurement. The stall is less abrupt, with a shallower post-stall $C_L$ slope. The lift coefficient levels off in the post-stall region, which is also observed in the WMLES results. Figure \ref{fig:ed1974_perf}(b) compares drag coefficients between the methods. At stall, both the clean and glaze ice geometries undergo a rapid rise in drag due to the onset of separation. The rapid decrease and subsequent increase in drag observed in the experimental wake drag measurements at high angles of attack is due to the experimental uncertainties \cite{broeren2018icingaeroperf_naca23012}. This discrepancy was not present in the force balance measurements. In Figure \ref{fig:ed1974_cp}, we look at spanwise averaged pressure coefficients compared to pressure probes on the glaze ice geometry at four angles of attack. In Figure \ref{fig:ed1974_cp}(a-b), reasonable agreement is observed between the experimental and WMLES results leading to less than 10\% error in the lift coefficient results. These two cases represent a pre-stall angle, $\alpha=6\degree$, and  $\alpha=10\degree$ corresponding to $C_{L,max}$. At higher angles of attack [Figure \ref{fig:ed1974_cp}(c-d)], there is an over-prediction of lift generation near the leading edge of the suction side of the airfoil. For $\alpha=12\degree$, the $C_p$ values are both underpredicted and overpredicted in different regions leading to the cancellation of errors, and therefore a good agreement in the lift coefficient can be seen. Both cases result in a delayed region of adverse pressure gradient near the leading edge of the airfoil as compared to the experimental result. For $\alpha=14\degree$, the pressure recovers to the experimental value around $x/c=0.5$. A small pocket of lift is seen near $x/c=0.85$ for Figure \ref{fig:ed1974_cp}(d). Additionally, the moment coefficients are underpredicted in the post-stall region compared to the experimental results. These discrepancies are due to non-physical domain constraints and are discussed in detail in Section \ref{sec:span}. Generally, we find good agreement between our WMLES calculations and the experimental results.

\subsubsection{Spanwise variation}\label{sec:span}

\begin{figure}
    \centering
    \includegraphics[width=0.75\textwidth]{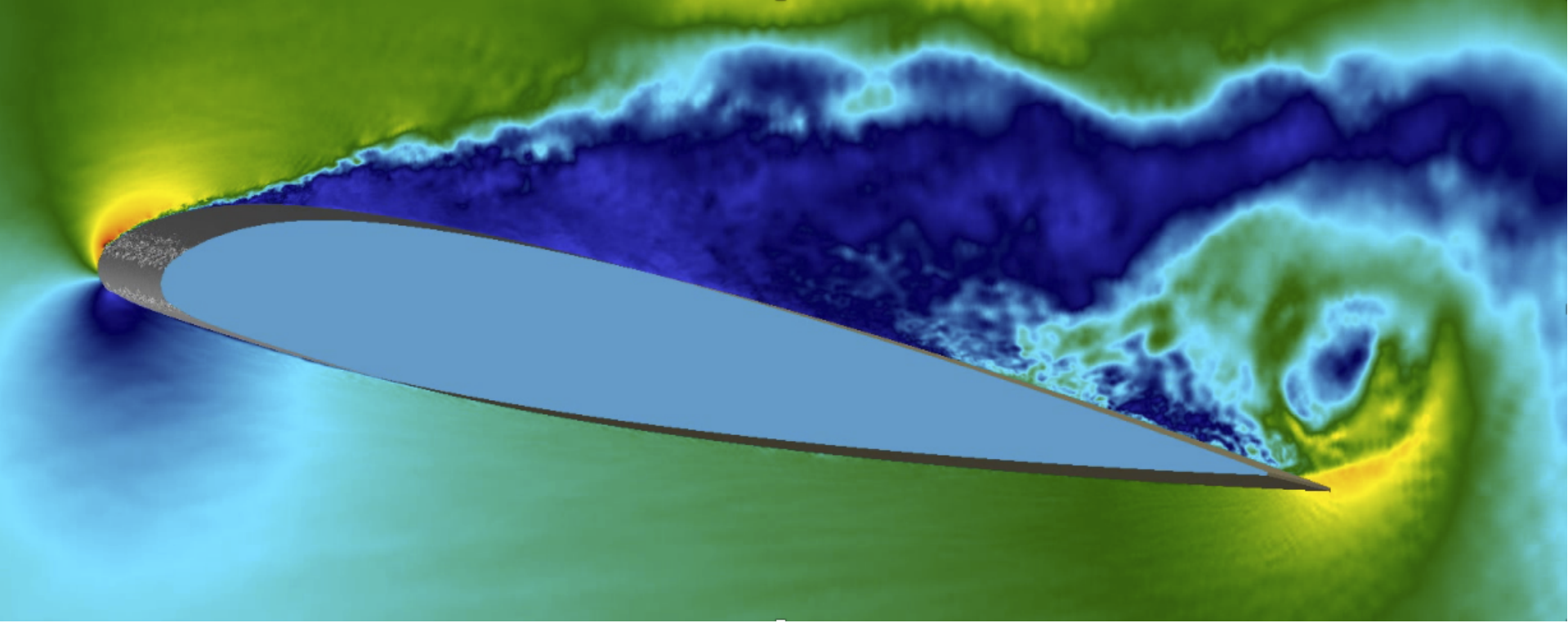}
    \caption{Instantaneous center-line flow-field snapshot of velocity magnitude during a sixteen-degree angle of attack ($\alpha=16^{\degree}$) simulation.}
    \label{fig:ed1974_stall_snap}
\end{figure}

If we look at a snapshot of the resulting flow field (see Figure \ref{fig:ed1974_stall_snap}), we observe two-dimensional vortical structures. These are believed to be caused by the spanwise restriction imposed by using a periodic domain with a $0.2c$ extent. To investigate this further, we conduct a spanwise extent study for early-time glaze ice case at $\alpha=9.3^{\degree}, 14^{\degree}$, and $16^{\degree}$. Table \ref{tab:grid_description} shows the details of the grids for the spanwise study. Three additional simulations are performed with a span of $0.4c$, $0.8c$, and $1.6c$.

\begin{figure}
    \centering
    \includegraphics[width=0.55\textwidth]{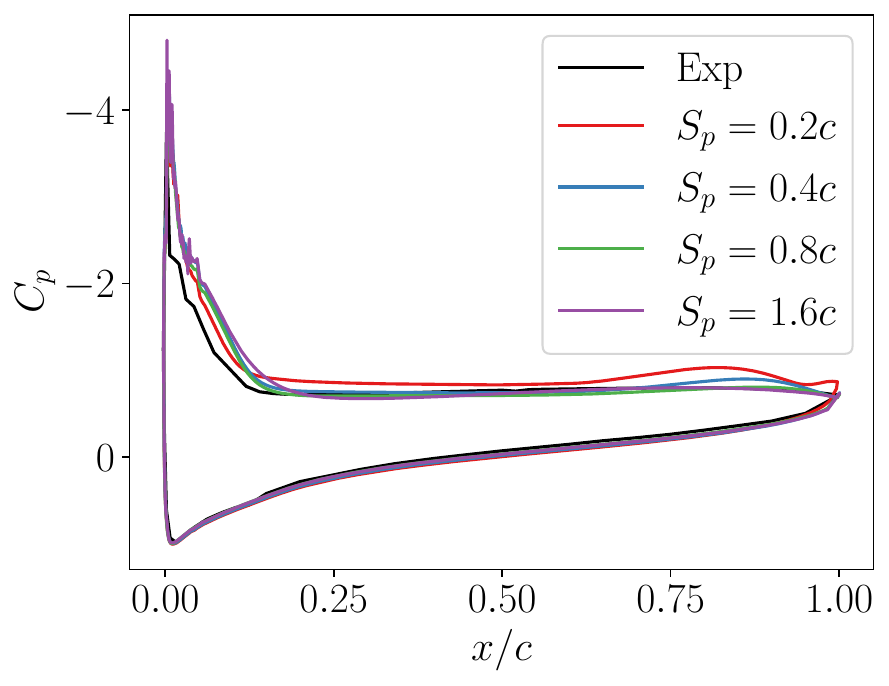}
    \caption{Sensitivity of pressure coefficients ($C_P$) as a function of $x/c$ with increasing span with comparisons to experimental pressure measurements \citep{broeren2018icingaeroperf_naca23012}.}
    \label{fig:cp_spanwise}
\end{figure}

\begin{figure}
    \centering
    \includegraphics[width=1.0\textwidth]{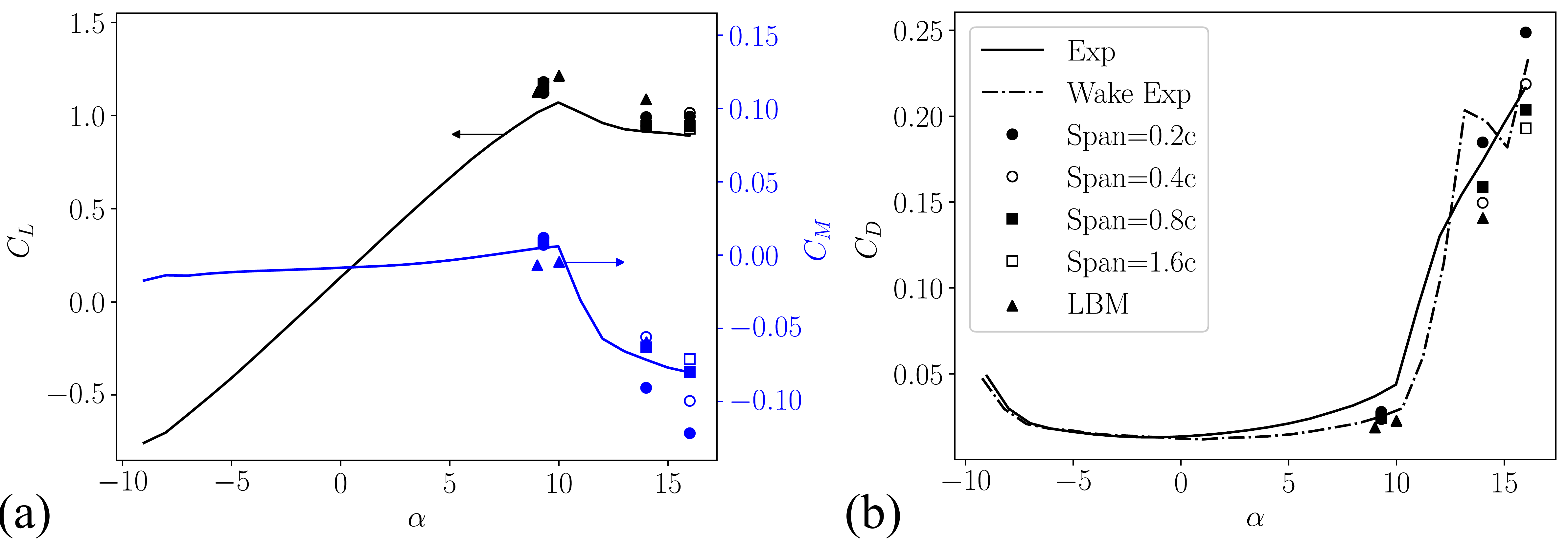}

    \caption{Sensitivity of (a) lift ($C_L$), moment ($C_M$), and (b) drag ($C_D$) coefficients for the glaze ice geometry with increasing span with comparisons to experimental wake (\chain) and force balance (\full) measurements \citep{broeren2018icingaeroperf_naca23012} as well as LBM simulations \citep{konig2015icelbm}. The legend applies to both panels.}
    \label{fig:perf_spanwise}
\end{figure}

In Figure \ref{fig:cp_spanwise}, spanwise conditioned pressure coefficients are plotted as a function of $x/c$, where $x$ is the spatial location along the chord. Minimal changes are observed near the leading edge ($x/c<0.2$). Around $x/c=0.85$, a small pocket of lower $C_p$ is observed. This region is indicative of a region of lift that is far from the center of the quarter-chord moment. As the span is increased, this pocket becomes small, indicating that it is an artifact of the spanwise restricted flow field for high angles of attack. Indeed, in Figure \ref{fig:perf_spanwise}(a), we observe drastic changes in the moment coefficients for high angles of attack. Increasing the span reduces the magnitude of the nose-down pitching moment and brings it into agreement with the experimental results. In contrast, as the span is increased, minimal changes are observed in the lift coefficients (Figure \ref{fig:perf_spanwise}(a)). Figure \ref{fig:perf_spanwise}(b) compares the drag coefficient for each span. At $\alpha=9.3^{\degree}$, minor sensitivity to the spanwise extent is observed. At higher angles of attack, the drag decreases with an increase in span. When simulating post-stall angles of attack, selecting a spanwise extent that minimizes errors due to the spanwise domain restrictions is necessary. This is especially necessary for predicting the moment coefficients accurately.

\subsection{Horn ice geometry}\label{sec:horn}
\begin{figure}
    \includegraphics[width=0.99\textwidth]{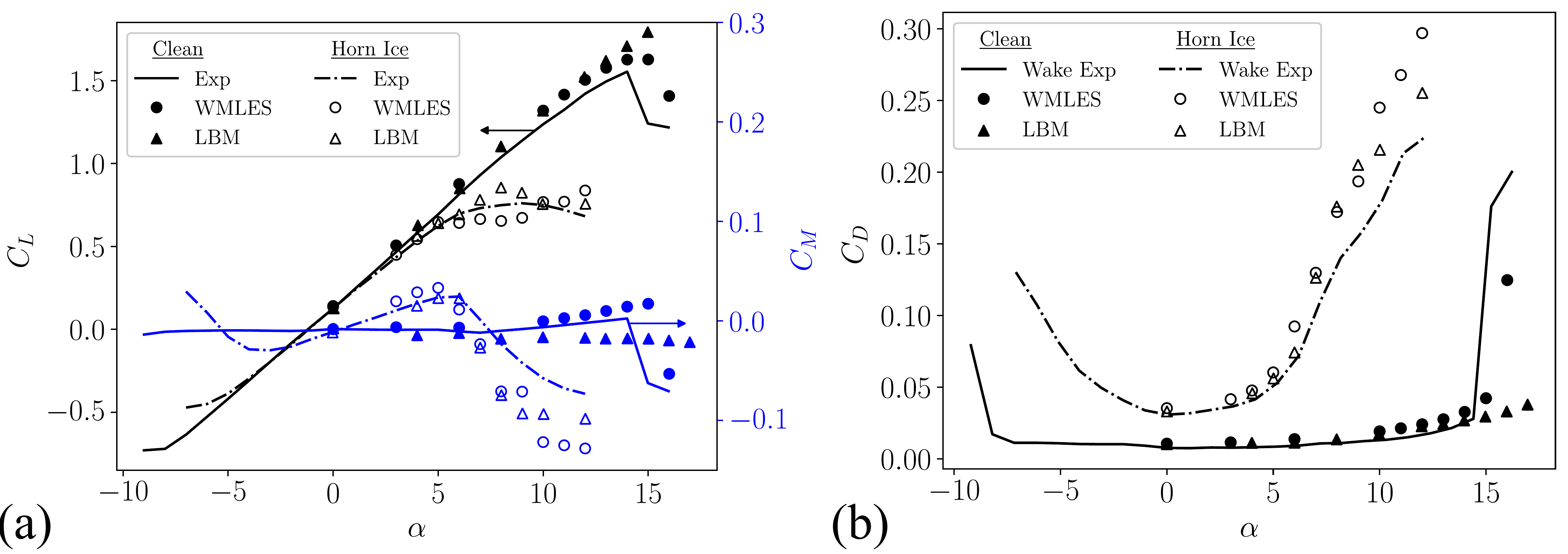}

    \caption{(a) Lift ($C_L$), moment ($C_M$), and (b) drag ($C_D$) coefficients comparing the clean (closed symbols, \full) and horn ice (open symbols, \chain) geometries of the experimental \citep{broeren2018icingaeroperf_naca23012}, LBM \citep{konig2015icelbm}, and present WMLES results.}
    \label{fig:horn_ice}
\end{figure}

\begin{figure}
    \includegraphics[width=0.99\textwidth]{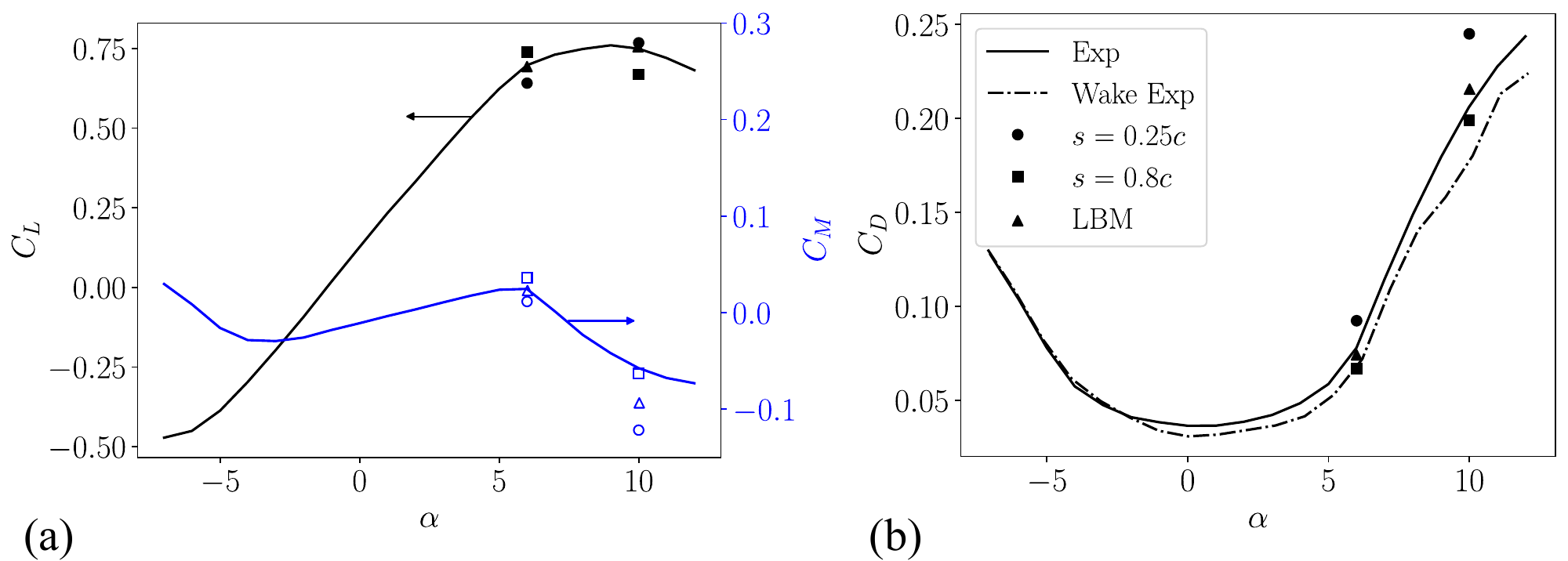}

    \caption{Sensitivity of (a) lift ($C_L$), moment ($C_M$), and (b) drag ($C_D$) coefficients for the horn ice geometry with increasing span with comparisons to experimental wake (\chain) and force balance (\full) measurements \citep{broeren2018icingaeroperf_naca23012} as well as LBM simulations \citep{konig2015icelbm}. The legend applies to both panels.}
    \label{fig:horn_span}
\end{figure}

Next, we assess the LES approaches for the horn ice geometry, shown in Figure \ref{fig:ice performance hit}(e). In Figure \ref{fig:horn_ice}, the simulated results' lift, moment, and drag coefficients are compared with the experimental and LBM results. We observe good agreement with the experimental data for the horn ice case. Similar results are obtained with the LBM approach. At high angles of attack, we observe overpredicted drag and underpredicted moments. Similar to Section \ref{sec:glaze}, we run additional cases with higher spanwise extents, here $S_p=0.8c$. In Figure \ref{fig:horn_span}, aerodynamic coefficients are assessed with respect to increasing span. Similar to the glaze ice case, increasing the spanwise extent improves the prediction of $C_M$, reduces $C_D$ (here, closer to the wake drag measurements, and has less effect on $C_L$). Increasing the span of the horn ice shape also incorporates more of the three-dimensional features of the original laser-scanned airfoil (see, Figure \ref{fig:pdf_rough}). The introduction of the additional scales with the increase in span can also affect drag and bring it closer to the experimental drag measurements (Figure \ref{fig:horn_span}(b)). 

\begin{figure}
    \centering
    \includegraphics[width=0.99\textwidth]{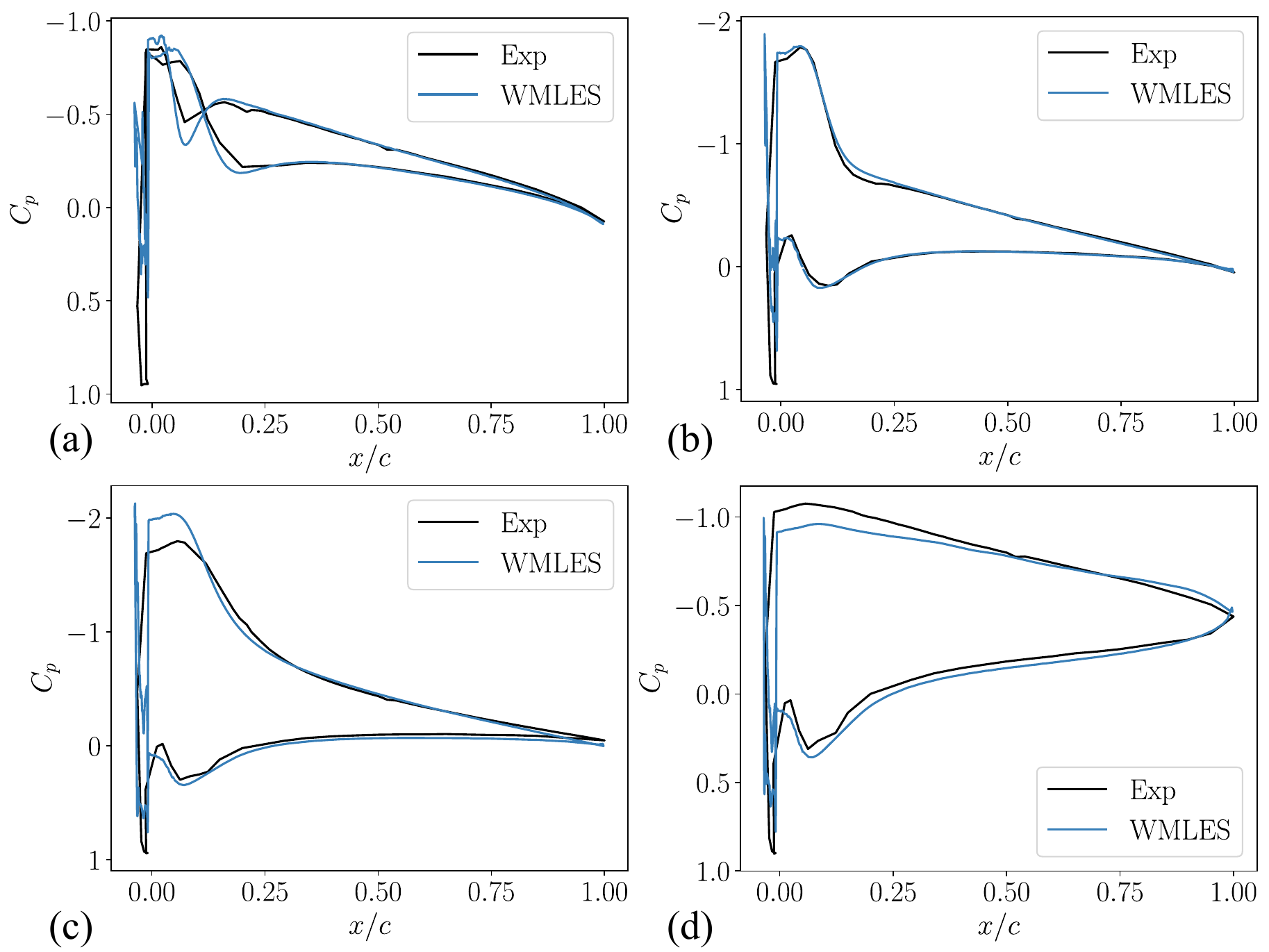}
    \caption{Pressure coefficient ($C_p$) comparison to experimental data for the horn ice geometry with angles of attack (and span) at (a) $0\degree$ ($S_p=0.25c$), (b) $4\degree$ ($S_p=0.25c$), (c) $6\degree$ ($S_p=0.8c$), and (d) $10\degree$ ($S_p=0.8c$) \citep{broeren2018icingaeroperf_naca23012}.}
    \label{fig:ed1978_cp}
\end{figure}

Pressure coefficients are compared to the experimental data in Figure \ref{fig:ed1978_cp} for two angles at pre-stall conditions in Figure \ref{fig:ed1978_cp}(a-b), one at $C_{L,max}$ in Figure \ref{fig:ed1978_cp}(c), and one at post-stall in Figure \ref{fig:ed1978_cp}(d). The spanwise extent is $0.25c$ for (a,b) and $0.8c$ for (c,d). Good agreement is observed across the cases. But at maximum lift, the region just downstream of the horn ice accretion is overpredicted, indicating a need for additional refinement or the inclusion of geometric features outside the chosen span.



\begin{figure}
    \centering
    \includegraphics[width=0.95\textwidth]{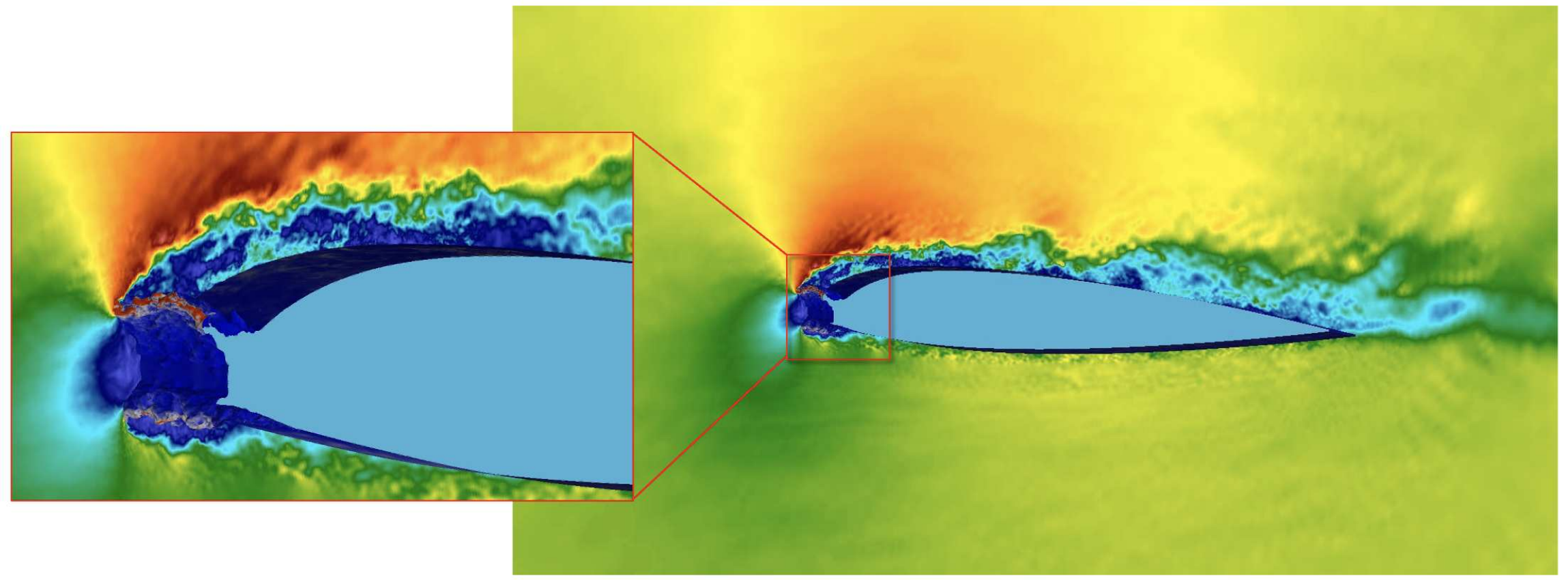}
    \caption{Center-line flow-field slice colored by velocity magnitudes with surface plot colored by wall shear stress of the NACA23012 airfoil at $\alpha=5^{\degree}$ with horn ice. Inset focuses on the leading edge of the airfoil, highlighting the high shear stress near the tips of each horn.}
    \label{fig:horn_ice_image}
\end{figure}

The horn ice case is less challenging for WMLES simulations, but more challenging for constructing body-fitted meshes due to the complex geometric features. In Figure \ref{fig:horn_ice_image}, an instantaneous center-line slice colored by the velocity magnitudes shows how the flow is immediately separated downstream of the horn ice shape even for a moderate angle of attack ($\alpha=5^{\degree}$). Wall shear stress contours on the surface highlight the localized value of shear stress at the tips of the ice horns. Flow separation occurs due to the geometric obstruction of the horns, whereas the other geometries, with less ice accretion, have less intrusive separation mechanisms.

\subsection{Early-time rime ice geometry}\label{sec:rime}

The third geometry we simulated is the early-time rime ice geometry, shown in Figure \ref{fig:ice performance hit}(b). 
Unlike the glaze ice geometry, the roughness is distributed evenly across the upper and lower surfaces of the NACA23012 airfoil, see Figure \ref{fig:roughness} (c). We apply the established practices from the glaze ice simulations (Section \ref{sec:glaze}) to the rime ice geometry. To the best of our knowledge, these are the first-ever simulations of a laser-scanned rime ice geometry.

\begin{figure}
    \centering
    \includegraphics[width=1.0\textwidth]{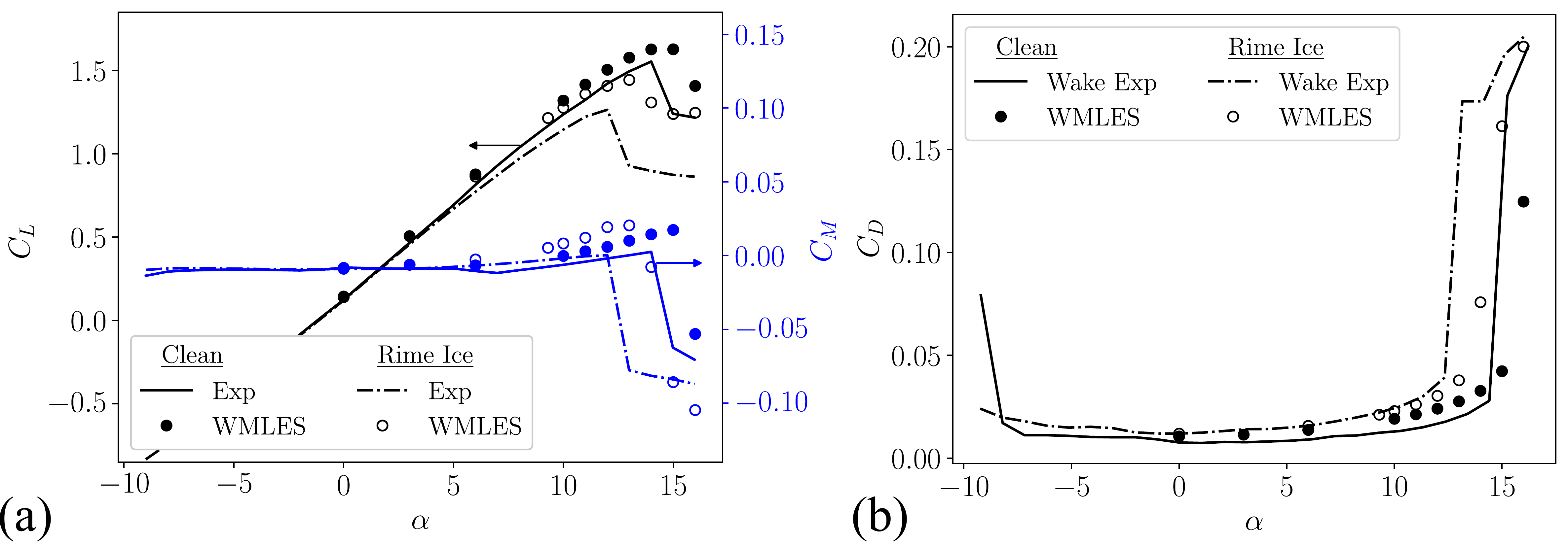}

    \caption{(a) Lift ($C_L$), moment ($C_M$), and (b) drag ($C_D$) coefficients comparing the clean (closed symbols, \full) and rime ice (open symbols, \chain) geometries of the experimental \citep{broeren2018icingaeroperf_naca23012} and present WMLES results.}
    \label{fig:rime_ice}
\end{figure}

In Figure \ref{fig:rime_ice}, lift, drag, and moment coefficients are compared with the experimental data. Contrary to the glaze ice case, the rime ice case does not predict the correct change in lift and drag. The maximum lift and critical stall angle are both overpredicted. The delayed separation can also be observed when comparing the wake drag measurements to the simulated results. These mispredictions are due to the under-resolved roughness scales near the leading edge. The maximum roughness height, as a percentage of chord length, is approximately $0.03\%$ for the rime ice geometry. This results in about $1.5$ control volumes per maximum roughness height; hence, all the roughness scales are in the sub-grid scale regime. 

\begin{figure}
    \centering
    \includegraphics[width=0.99\textwidth]{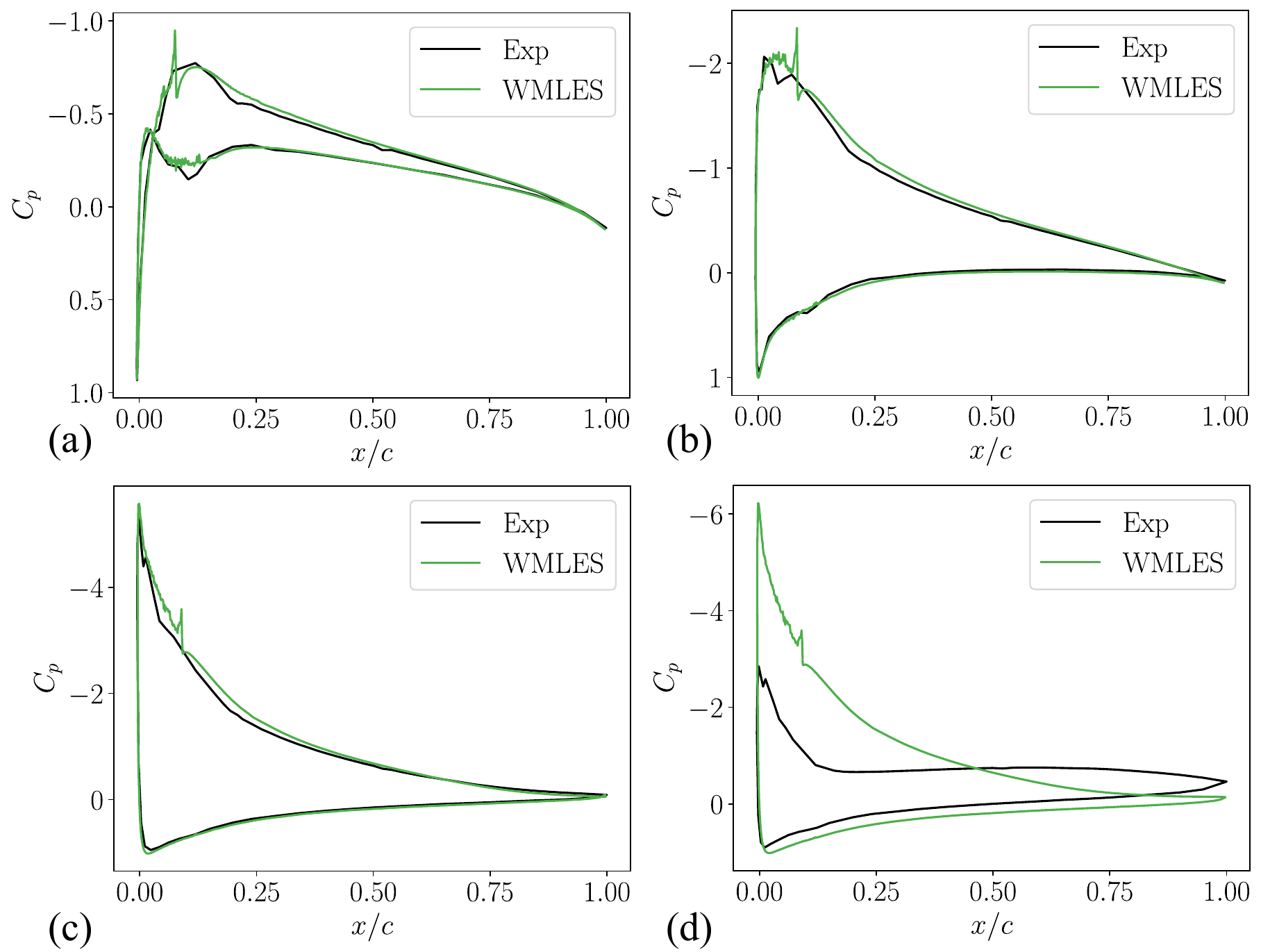}
    \caption{Pressure coefficient ($C_p$) comparison to experimental data for the rime ice geometry with angles of attack at (a) $0\degree$, (b) $6\degree$, (c) $12\degree$, and (d) $13\degree$ \citep{broeren2018icingaeroperf_naca23012}.}
    \label{fig:rime_cp}
\end{figure}

This is further emphasized in Figure \ref{fig:rime_cp}, where pressure coefficients are compared to the experimental data for two angles at pre-stall conditions in Figure \ref{fig:rime_cp}(a-b), one at $C_{L,max}$ in Figure \ref{fig:rime_cp}(c), and one at post-stall in Figure \ref{fig:rime_cp}(d). Here, minor overpredictions of $C_p$ lead to larger overshoots for $C_L$ (a-c). For the post-stall angle, we observe a large overprediction of $C_p$ on the suction side of the airfoil near the leading edge. A small separation event occurs near the trailing edge of the airfoil. For this angle of attack, the experimental observation is that the flow over the airfoil is fully separated.

\begin{figure}
    \centering
    \includegraphics[width=1.0\textwidth]{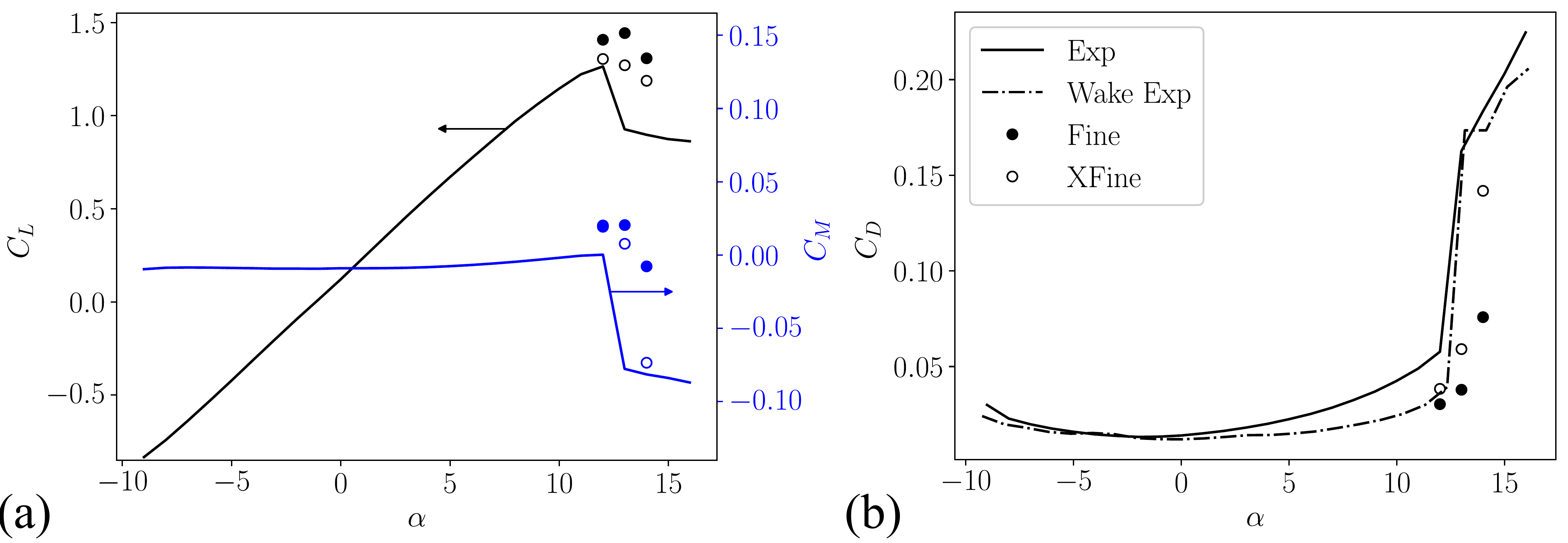}

    \caption{Sensitivity of (a) lift ($C_L$), moment ($C_M$), and (b) drag ($C_D$) coefficients for the rime ice geometry with increasing grid resolution with comparisons to experimental wake (\chain) and force balance (\full) measurements \citep{broeren2018icingaeroperf_naca23012}. The legend applies to both panels.}
    \label{fig:rime_refined}
\end{figure}

\begin{figure}
    \centering
    \includegraphics[width=1.0\textwidth]{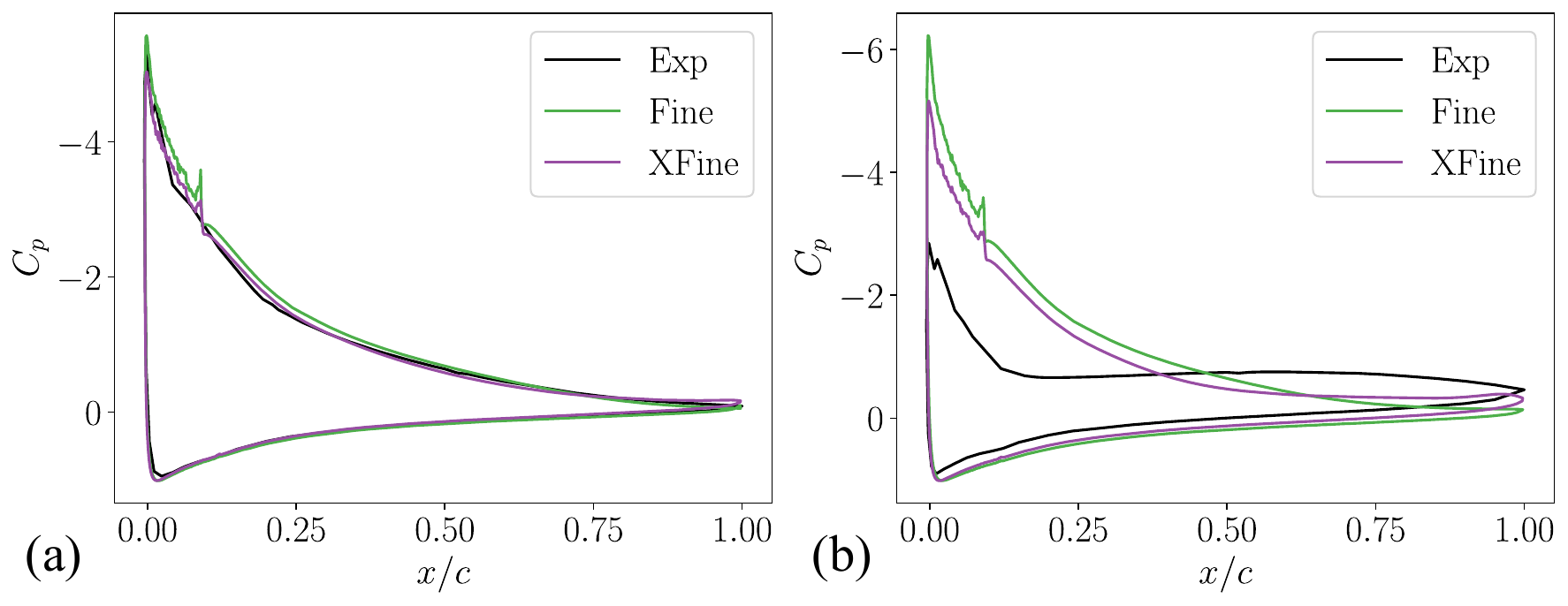}

    \caption{Sensitivity of pressure coefficients ($C_p$) for the rime ice geometry with increasing grid resolution with comparisons to experimental pressure measurements with angles of attack at (a) $12\degree$ and (b) $13\degree$ \citep{broeren2018icingaeroperf_naca23012}.}
    \label{fig:cp_rime_refined}
\end{figure}

In Figure \ref{fig:rime_refined}, we refine the grid to an additional level (equivalent to the extra fine clean airfoil simulation) for three angles of attack (one at $C_{L,max}$ and  two in the post-stall region).
It can be observed that predictions of the lift, drag, and moment coefficients improve with respect to the experimental data. Additionally, in Figure \ref{fig:cp_rime_refined}, we plot pressure coefficients for $\alpha = 12\degree$ (a) and $\alpha=13\degree$ (a) compared to experimental pressure coefficients for two grid resolutions (fine and extra fine). At the critical angle of attack, small reduction in $C_p$, with grid refinement, at the leading edge lead to a better prediction of $C_{L,max}$ (as seen in Figure \ref{fig:rime_refined}). In the post-stall angles, we observe only minor improved results as compared to the experimental data, but it is clear that additional refinement is required to accurately capture the stall event. 

There are approximately $3$ control volumes per maximum roughness height at this extra-fine resolution. For the methods used in this work, the recommended number of points per roughness height is at least $4$ \citep{goc2022ctr, joo2016roughness} to resolve the roughness features. Here, we are beginning to resolve the roughness scales, but further refinement is required to resolve all the scales appropriately. These refinement regimes begin to approach wall-resolved resolutions, limiting their utility in engineering applications. 

\begin{figure}
    \centering
    \includegraphics[width=1.0\textwidth]{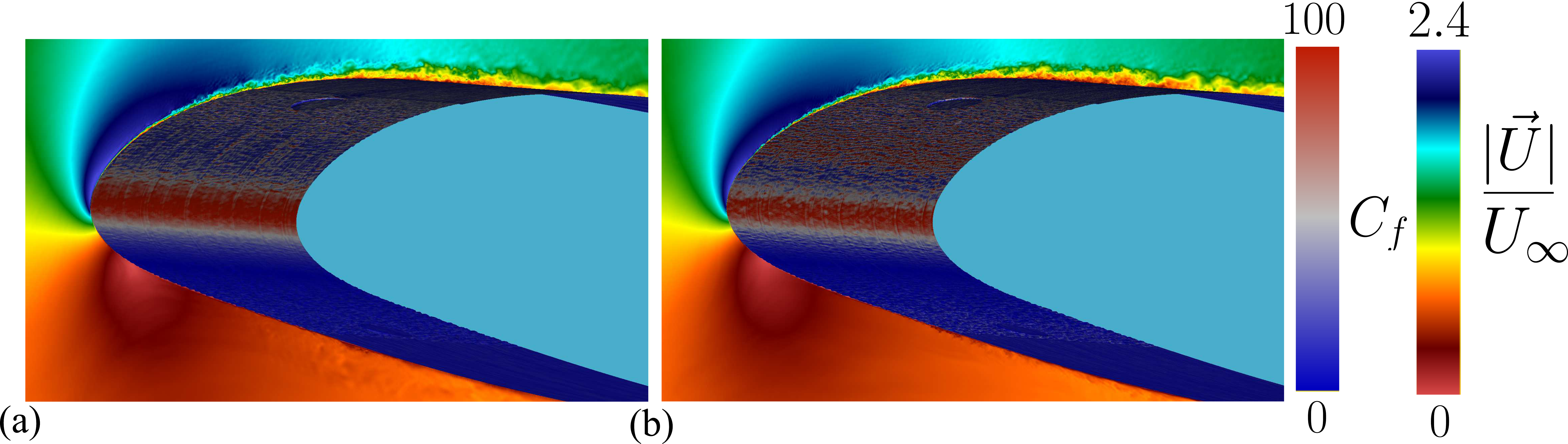}

    \caption{Center-line flow-field slice colored by velocity magnitudes with surface plot colored by wall friction coefficient of (a) fine and (b) extra-fine rime ice geometry at $\alpha=12^{\degree}$.}
    \label{fig:rime_slice_ref}
\end{figure}

In Figure \ref{fig:rime_slice_ref}, we compare the qualitative features of the flow field for the fine case [Figure \ref{fig:rime_slice_ref}(a)] and the extra fine case [Figure \ref{fig:rime_slice_ref}(b)]. Comparing the two results highlights that the roughness elements are completely sub-grid for the fine-grid simulations. For the fine case, we observe a flow field qualitatively similar to those observed in the clean geometry. In contrast, for the extra fine case, regions of increased wall shear stress can be observed in the regions with rime ice roughness. The additional refinement required to capture the roughness elements becomes intractable when simulating a wing or complete aircraft under similar resolutions. Therefore, these results highlight the need for roughness wall models in rime ice conditions.


\subsection{Streamwise ice geometry}

The streamwise ice geometry, shown in Figure \ref{fig:ice performance hit}(d), is the resulting shape from a five-minute exposure to rime ice conditions (see Table \ref{tab:ice_description}). Small droplets accrete at the leading edge geometry and freeze before any run back can occur. This freezing process leads to so-called spear-like geometries where the blunt leading edge of the airfoil becomes sharper. Similar to the early-time rime ice shape, the roughness elements are distributed more uniformly across the leading edge of the airfoil (Figure \ref{fig:stats_rime}). The roughness scales reach 2-4 times the size of the early-time rime ice geometry. This increases the number of grid points across a given roughness element. Small roughness elements close to the leading edge will still be highly under-resolved.

\begin{figure}
    \centering
    \includegraphics[width=1.0\textwidth]{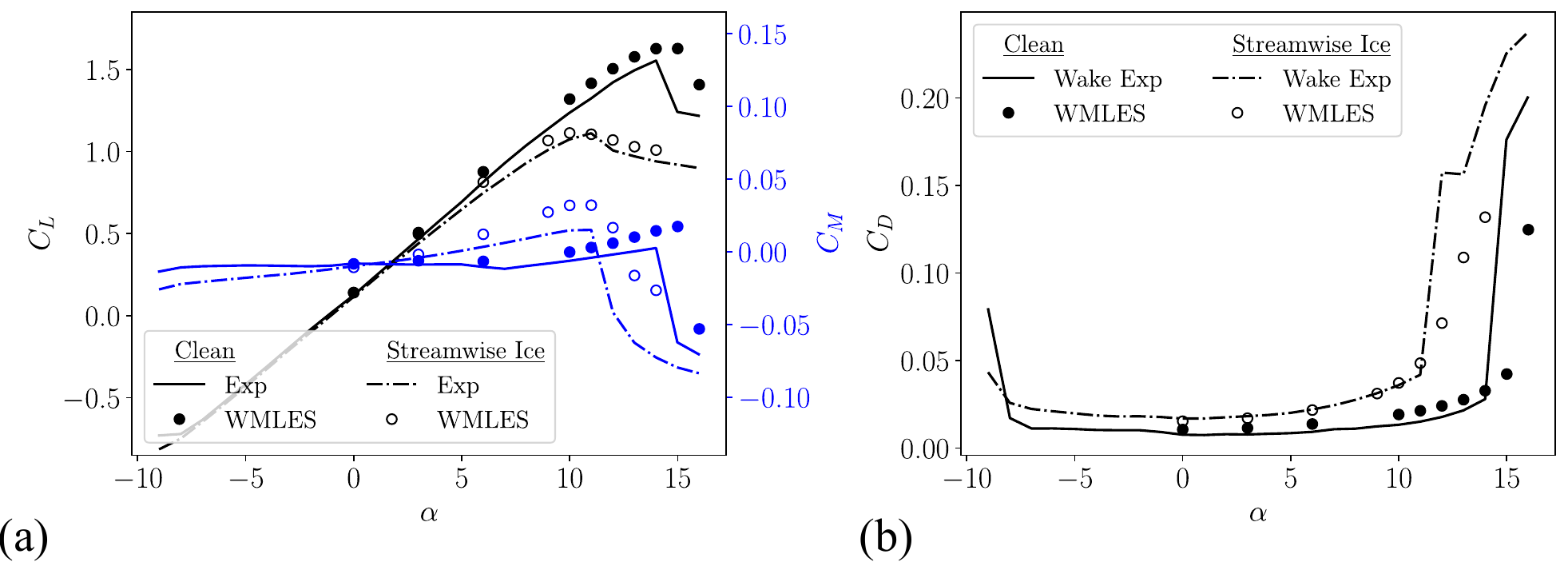}
    \caption{(a) Lift ($C_L$), moment ($C_M$), and (b) drag ($C_D$) coefficients comparing the clean (closed symbols, \full) and streamwise ice (open symbols, \chain) geometries of the experimental \citep{broeren2018icingaeroperf_naca23012} and present WMLES results.}
    \label{fig:ed1977_perf}
\end{figure}

\begin{figure}
    \centering
    \includegraphics[width=0.99\textwidth]{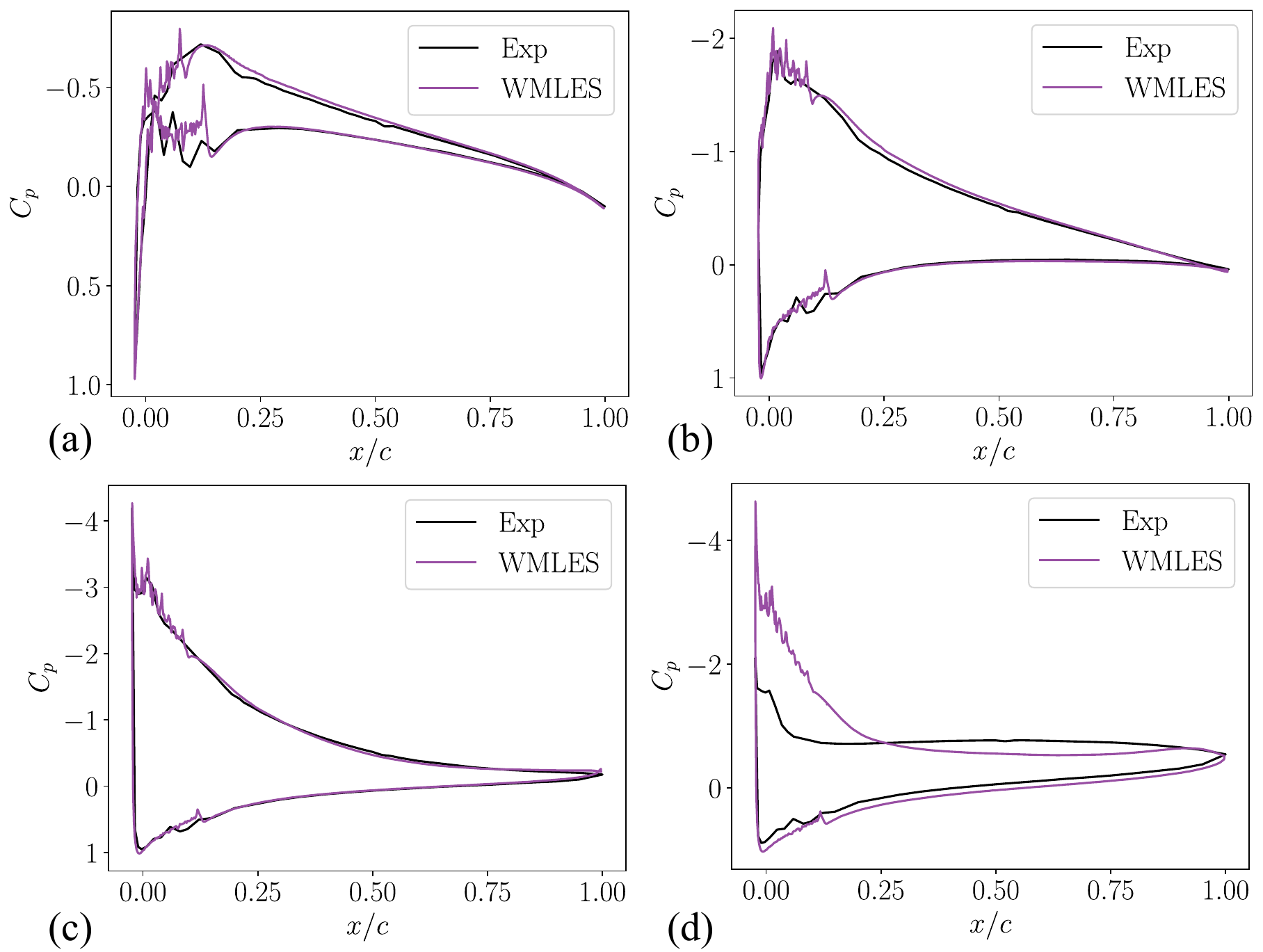}
    \caption{Pressure coefficient ($C_p$) comparison to experimental data for the streamwise ice geometry with angles of attack at (a) $0\degree$, (b) $6\degree$, (c) $11\degree$, and (d) $14\degree$ \citep{broeren2018icingaeroperf_naca23012}.}
    \label{fig:ed1977_cp}
\end{figure}

In Figure \ref{fig:ed1977_perf}, lift, drag, and moment coefficients are compared to the experimental data. In the region prior to stall (here $\alpha\approx 11^\circ$), both lift and drag coefficients have good agreement with the experimental results. In the post-stall region, the lift coefficient follows the downward slope of the experiment, which has a less abrupt stall behavior than the rime ice geometry. While the chosen grid resolution resolves the macroscopic effects of lift and drag (i.e., critical stall angle and $C_{L,max}$), it is still not sufficient to resolve many of the roughness scales which results in lower drag coefficient for post-stall angles. Pressure coefficients at four angles of attack are shown in Figure \ref{fig:ed1977_cp}. We observe good agreement with the experimental measurements up to $\alpha=11^\circ$, which is representative of the location of $C_{L,max}$ for these conditions. In Figure \ref{fig:ed1977_cp}(d), cancellation of errors between the leading edge and trailing edge pressure distribution explain our reasonable agreement with $C_L$, but overprediction of $C_M$ in Figure \ref{fig:ed1977_perf}(a).

\subsection{Spanwise ridge ice geometry}\label{sec:spanwise_ridge}


The final ice shape we consider is the spanwise ridge geometry. This geometry mimics the use of a thermal-based icing protection system (IPS) for the leading edge of a wing. Specifically, the icing conditions and heater settings were determined to induce run-back of the water past the IPS. The inclusion of the IPS makes this geometry unique from the other four. This can be seen in Figure \ref{fig:ice performance hit}(g), which shows the outer mold line of the spanwise ridge shape. The leading edge is identical to the clean NACA23012 geometry, and at some distance downstream, just past the IPS, a prominent ridge of ice is formed \cite{broeren2015nasatr}. For this $Re_c$, the primary adverse effect of the ridge is a significant increase in drag. The maximum lift coefficient differs by only 3-tenths of a lift coefficient compared to the clean geometry. Additionally, the stall is delayed by approximately $2^\circ$. According to \cite{broeren2015nasatr}, at increased $Re_c$, the spanwise ridge shape would result in much lower values of $\alpha_{crit}$ and $C_{L,max}$. 

\begin{figure}
    \centering
    \includegraphics[width=1.0\textwidth]{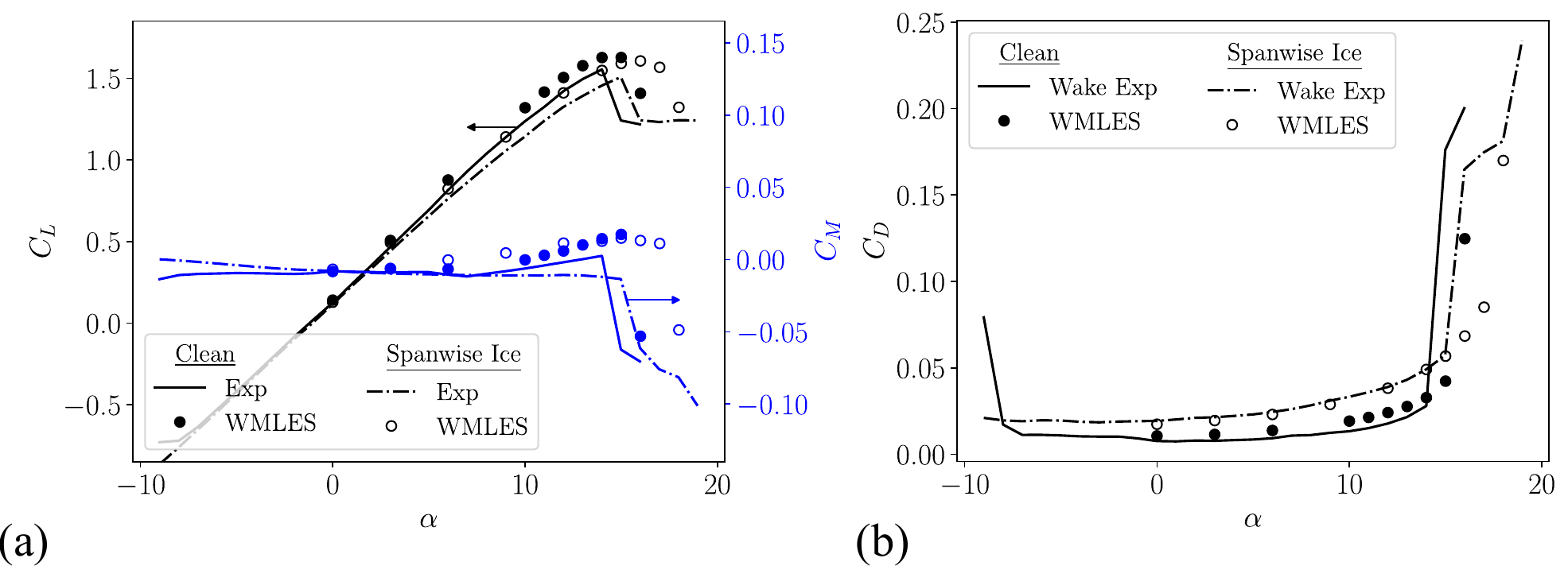}
    \caption{(a) Lift ($C_L$), moment ($C_M$), and (b) drag ($C_D$) coefficients comparing the clean (closed symbols, \full) and spanwise ridge ice (open symbols, \chain) geometries of the experimental \citep{broeren2018icingaeroperf_naca23012} and present WMLES results.}
    \label{fig:ed1967_perf}
\end{figure}

\begin{figure}
    \centering
    \includegraphics[width=0.99\textwidth]{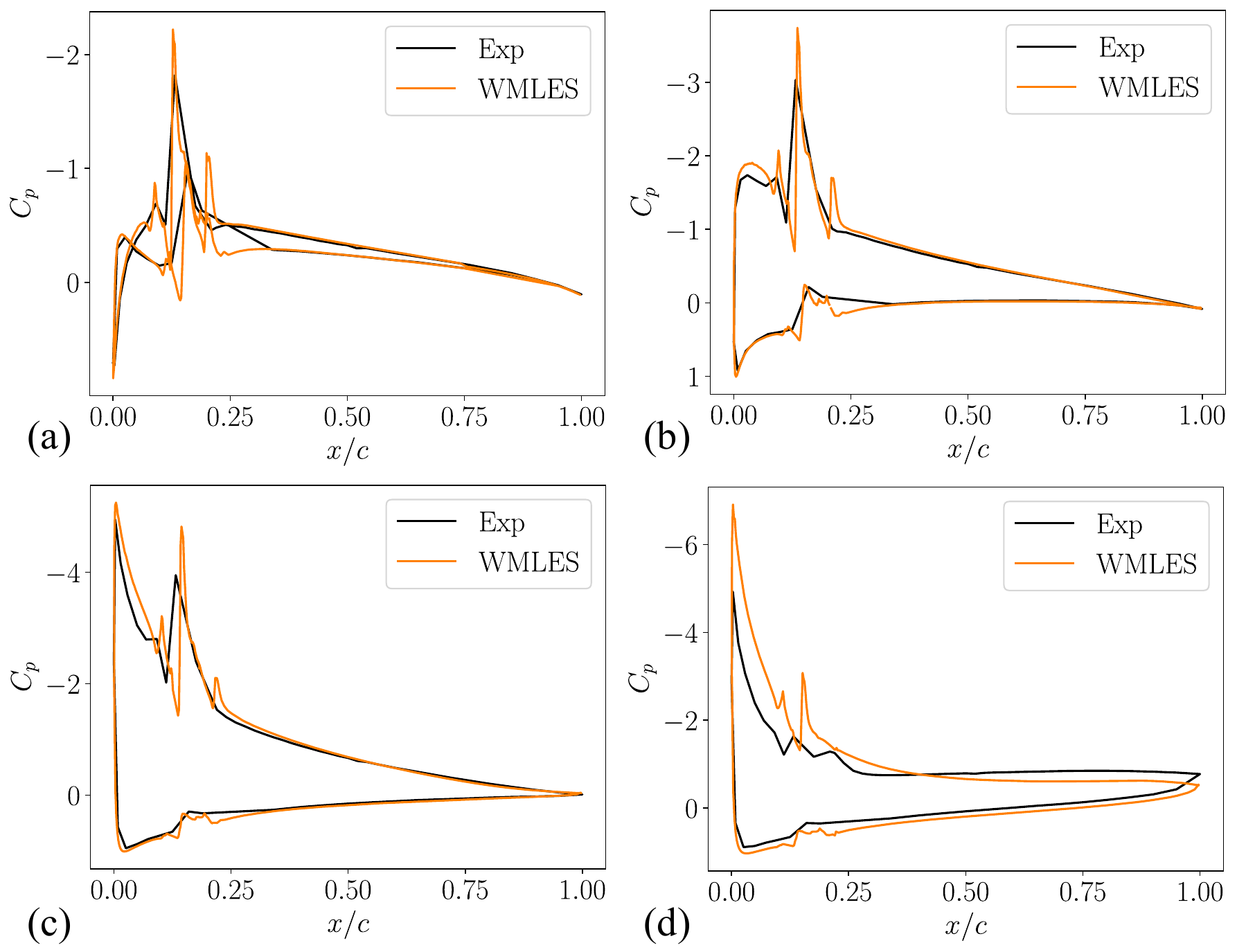}
    \caption{Pressure coefficient ($C_p$) comparison to experimental data for the spanwise ridge ice geometry with angles of attack at (a) $0\degree$, (b) $6\degree$, (c) $12\degree$, and (d) $18\degree$ \citep{broeren2018icingaeroperf_naca23012}.}
    \label{fig:ed1967_cp}
\end{figure}

In Figure \ref{fig:ed1967_perf}, we compare the integrated aerodynamic coefficients to the experimental data. We observe a delayed stall (higher critical $\alpha$) for the spanwise ridge ice shape, like the experimental results. In contrast to the experiment, the critical angle of attack in the simulations occurs at an $\alpha$ value approximately $1^\circ-2^\circ$ larger than the experimental result. This shift is also observed in $C_M$. We overpredict the critical angle of attack for the fully clean geometry by approximately $1^\circ-2^\circ$ as well. For both the clean and spanwise ice shape, the airfoil's leading edge is the same. In this region, we encounter turbulence transition and high-pressure gradients that stress the assumptions of the equilibrium wall model. Figure \ref{fig:ed1967_perf}(b) shows the increased drag due to the ice shape. We observe good agreement before stall with the wake drag measurements. The agreement with the wake drag in the post-stall region ($\alpha\approx18^\circ$) is reasonable. Pressure coefficients for $\alpha=0^\circ, 6^\circ, 12^\circ$, and $18^\circ$ are plotted in Figure \ref{fig:ed1967_cp}. Before stall, we find good agreement between the experimental and WMLES results. At the location of the ridge, an increase in pressure is observed, followed by a geometrically induced flow separation just downstream of the ridge line. This small separation bubble locally energizes the boundary layer, causing it to remain attached at higher angles than the fully clean case \cite{whalen2006considerations}.

\section{Conclusions}\label{sec:conclusion}

In this work, five NACA23012 ice geometries, along with a clean geometry, have been simulated using WMLES. These include early-time rime, streamwise (long exposure to rime), early-time glaze, horn (long exposure to glaze), and spanwise ridge (due to thermal protection system) ice shapes that represent the most critical ice shapes encountered in icing conditions \cite{bragg2005iced}. For each case, relevant comparisons are made to the available experimental and simulation results \cite{broeren2018icingaeroperf_naca23012, konig2015icelbm} focusing on both integrated and surface quantities, such as lift, drag, moment, pressure coefficients, and wall shear stress. Ice geometries show a bimodal behavior in the icing roughness where the characteristic ice height, $h$, could be separated from the local roughness height, $k$. This bimodal representation has implications for future modeling of rough iced surfaces.

The baseline NACA 23012 clean geometry is simulated for a series of grid resolutions. We observed a good agreement in the linear regime across various grids, but we required an increased resolution to simulate stall successfully. Additional resolution increases, leading to a grid with 54 million cells, yielded increased accuracy in the lift, drag, and moment coefficients. It was determined to utilize a grid resolution with 14.25 million cells as the primary resolution for the ice shapes as it is considered a tractable resolution for simulating more complicated aircraft-scale configurations of interest.

We observe improved results for the early-time glaze ice case compared to those established in the literature \cite{konig2015icelbm}. The WMLES results accurately capture the critical stall angle, observe the appropriate decrease in lift in the post-stall region, and yield good agreement with the rising drag coefficient at high angles of attack. 

A spanwise extent study was conducted for two reasons: (1) to eliminate non-physical vortices causing pockets of lift near the trailing edge and (2) to better represent the roughness statistics by including more of the original ice shape. We performed a spanwise extent study for three angles of attack ($\alpha=9.3\degree$, $\alpha=14\degree$, and $\alpha=16\degree$). Longer span resulted in improvements for both the lift and moment coefficients in the post-stall regions. Minimal differences were observed at $9.3\degree$. This highlights the need to simulate increased span at higher angles of attack to minimize any correlations in the flow caused by the reduced spanwise extent.

Similar results as compared to \cite{konig2015icelbm} were obtained for the horn ice geometry. It was shown that the increased spanwise extent was necessary to appropriately model the post-stall region. In contrast to the early-time glaze, larger differences are observed in the pre-stall angle of attack with an increased spanwise extent. This is primarily due to the inclusion of additional geometric, $h$, and roughness, $k$, length scales in the spanwise direction since the horn-ice geometry has more spanwise inhomogeneity compared to the early-time glaze ice shape. For both geometries, we observed good agreement in $C_p$ at select angles of attack.

For the early-time rime geometry, it was observed that a 21 million cell resolution was inadequate to accurately simulate the critical angle of attack and $C_{L,max}$. The roughness scales are much smaller for this geometry. For this grid resolution, all the roughness scales are sub-grid. Therefore, additional grid resolution was required to accurately capture both $\alpha_{crit}$ and $C_{L,max}$. With this additional refinement, improvements were observed in both $C_M$ and $C_D$ coefficients. The rime ice geometry is a good candidate case for future wall modeling of rough surfaces because of the additional resolution requirements.

Compared to the early-time rime ice shape, better results are obtained for the streamwise ice geometry. The roughness scales of this geometry are larger, and therefore, the same grid resolution results in a better representation of the roughness and how it affects the flow field. Reasonable agreement is observed for the $C_L$ and $C_M$ coefficients. The post-stall $C_D$ is underpredicted but is accurate in the linear regime pre-stall.

The last geometry discussed is the spanwise ridge ice shape, which mimics the use of an ice protection system on the leading edge of the airfoil. This results in a region of clean airfoil followed by a ridge shape located downstream of the leading edge. At the simulated $Re_c$, the primary adverse effect of the ridge is the increased drag, which is accurately captured by the WMLES at the pre-stall angles. The stall angle is delayed due to a small separation bubble that locally energizes the boundary layer. The WMLES results capture the delay in the critical stall angle but overpredict it by approximately 1\degree.

The relevant comparisons to the experimental results of \cite{broeren2018icingaeroperf_naca23012} show that qualitative and reasonable quantitative agreement with the experimental data is observed across all geometries. This is due to the combination of our modeling choices: Voronoi grid generation that can resolve complex roughness elements, low-dissipation numerics, advanced wall models, and dynamic sub-grid models. These modeling efforts provide heightened confidence in the application of WMLES for accurately predicting the complex effects of ice shapes on aerodynamic performance. Collectively, these findings mark a significant step towards bridging the existing modeling gap and advancing our understanding of complex ice shapes' impact on aerodynamic behavior and our ability to simulate them accurately.




\appendix

\section{Transition sensitivity to grid resolution}
\label{sec:app:grid_res}

\begin{figure}
    \centering
    \includegraphics[width=1.0\textwidth]{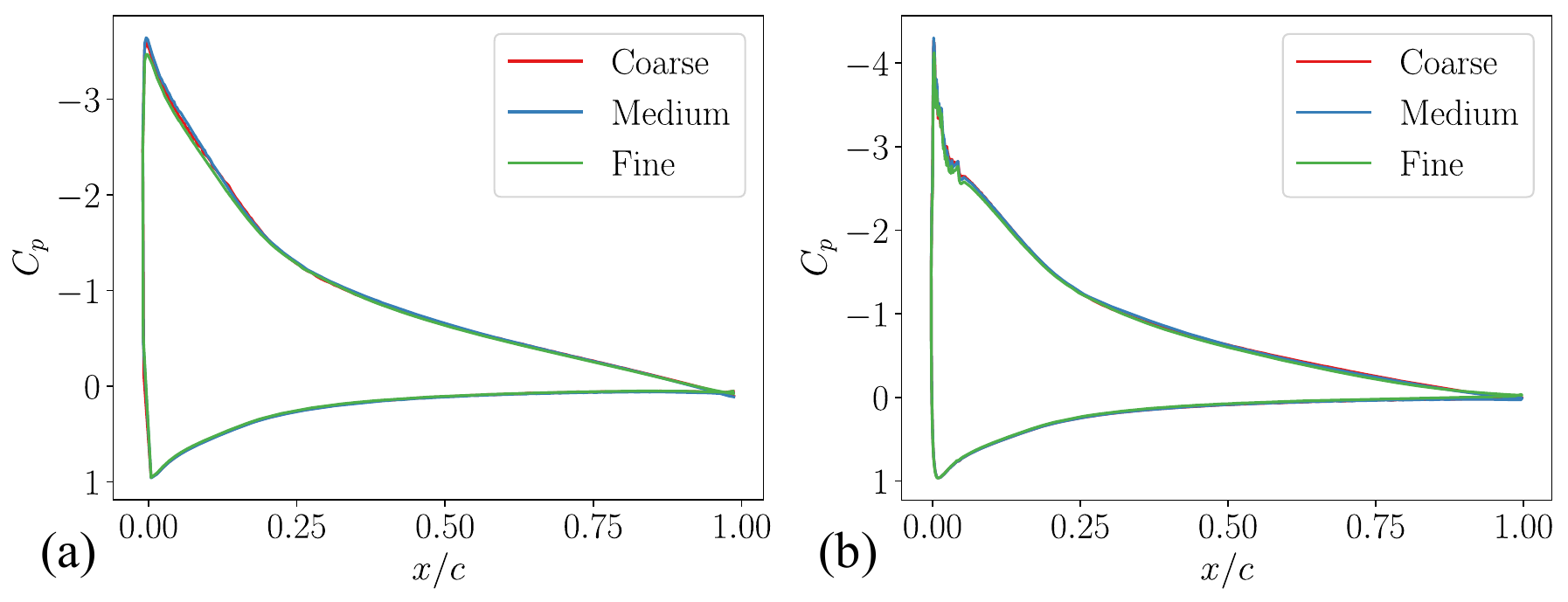}

    \caption{Sensitivity of pressure coefficients ($C_p$) for the (a) clean and (b) early-time glaze ice geometry with increasing grid resolution for $\alpha=9.3\degree$.}
    \label{fig:cp_clean_glaze_ref}
\end{figure}

\begin{figure}
    \centering
    \includegraphics[width=1.0\textwidth]{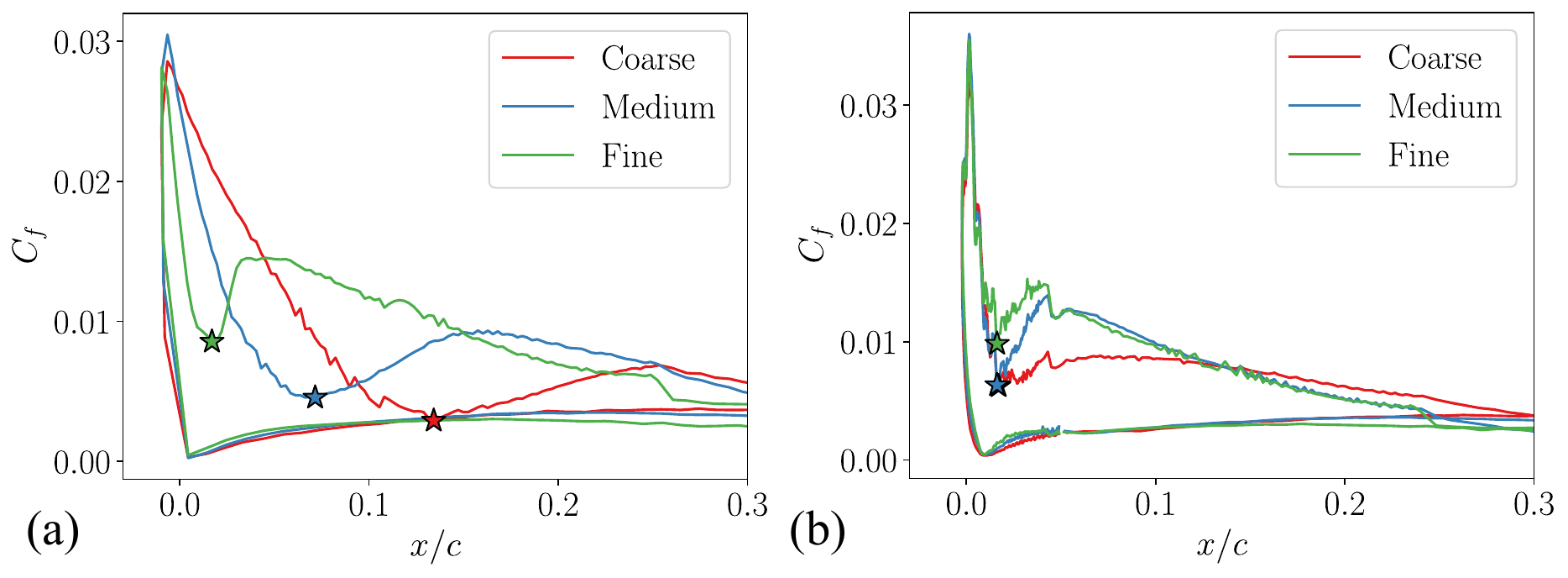}

    \caption{Sensitivity of friction coefficients ($C_f$) for the (a) clean and (b) early-time glaze ice geometry with increasing grid resolution for $\alpha=9.3\degree$. Flow transition is indicated by $\bigstar$ for each grid resolution.}
    \label{fig:tauw_clean_glaze}
\end{figure}

\begin{figure}
    \centering
    \includegraphics[width=0.6\textwidth]{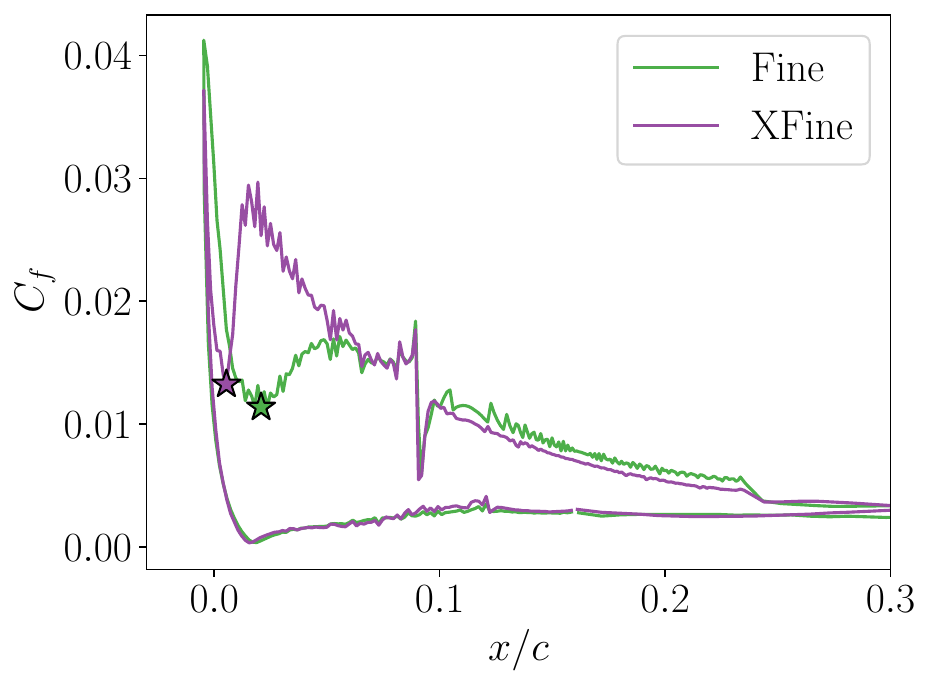}

    \caption{Sensitivity of friction coefficients ($C_f$) for the early-time rime ice geometry with fine and extra-fine grid resolutions at $\alpha=12\degree$. Flow transition is indicated by $\bigstar$ for each grid resolution.}
    \label{fig:tauw_rime}
\end{figure}

In this work, the grid resolution utilized in all simulations (denoted as fine), other than where expressly noted, uses a resolution representative of a high-lift Common Research Model (CRM-HL) configuration with a total grid count of 1.5 billion CV \cite{goc2023thesis}. To illustrate that this is an appropriate grid resolution, we conduct a grid resolution study for the early-time glaze ice configuration at an angle of attack of $9.3\degree$. We additionally compare the results for the glaze ice at these grid resolutions to those of the clean geometry. Figure \ref{fig:cp_clean_glaze_ref} compares pressure coefficients for three grids at coarse, medium, and fine resolutions. We observe reasonable grid convergence between the three resolutions for both the clean [Figure \ref{fig:cp_clean_glaze_ref}(a)] and early-time glaze ice [Figure \ref{fig:cp_clean_glaze_ref}(b)] geometries. As suggested by Figure \ref{fig:fig3_clean_perf}, additional refinement may be required to converge the integrated quantities at and beyond the stall angle. With the current modeling approach, the transition is handled implicitly, meaning we do not employ an additional transition model for the leading edge region. In Figure \ref{fig:tauw_clean_glaze}, we plot the friction coefficient, $C_f$, against the streamwise coordinate for the clean (a) and early-time glaze ice (b) conditions. We observe that the transition location is susceptible in the clean geometry, as increasing resolution leads to earlier transitions.
In contrast, the early-time glaze ice geometry shows minor sensitivity to transition location with increasing grid resolution. This illustrates that the roughness elements act as a tripping mechanism for the boundary layer. This is not always the case. For early-time rime ice geometries, the scale of the roughness elements is small enough that at our fine
grid resolutions, we do not have enough points per roughness height to capture this roughness-induced transition. This is observed in Figure \ref{fig:tauw_rime}, where increasing the grid resolution from fine to extra-fine moves the transition location closer to the leading edge. 

 \bibliographystyle{elsarticle-num} 
 \bibliography{cas-refs}





\end{document}